\documentclass[sigplan,10pt,nonacm]{acmart}

\microtypesetup{expansion=false}

\setcopyright{none}
\renewcommand\footnotetextcopyrightpermission[1]{}
\usepackage{graphicx}
\usepackage{multirow}
\usepackage{algorithm}
\usepackage{algorithmic}
\usepackage{subcaption}
\usepackage[skins]{tcolorbox}
\usepackage{float}
\IfFileExists{placeins.sty}{\usepackage{placeins}}{\newcommand{\FloatBarrier}{}}

\graphicspath{{figures/}}

\newcommand{\parabf}[1]{\vspace{0.05cm}\noindent\textbf{#1}}
\newcommand{\spork}{\textsc{Spork}}
\newcommand{\Ttool}{T_{\mathrm{tool}}}
\newcommand{\Tdec}{T_{\mathrm{dec}}}
\newcommand{\Tbase}{T_{\mathrm{base}}}
\newcommand{\Tbasestar}{T_{\mathrm{base}}^{*}}
\newcommand{\Toh}{T_{\mathrm{oh}}}
\newcommand{\toverlap}{t_{\mathrm{overlap}}}
\newcommand{\ftool}{f_{\mathrm{tool}}}
\definecolor{QuoteBlue}{HTML}{2A5487}
\newcommand{\introepigraph}[3]{%
  \begin{quote}
    \color{QuoteBlue}\itshape
    ``#1''\nobreak\hfill
    {\normalfont\small --- \textsc{#2}%
    \if\relax\detokenize{#3}\relax\else
      , \footnotesize\textit{#3}%
    \fi}
  \end{quote}
  \vspace{0.05em}
}

\title{SPORK: Self-Speculative Forking to Accelerate Agentic LLM Inference}
\author{Huajun Bai}
\authornote{Contact: baihuajun24@gmail.com}
\affiliation{%
  \institution{Tsinghua University}
  \country{}}

\author{Weiwei Lv}
\affiliation{%
  \institution{Meituan}
  \country{}}

\author{Huichuan Zheng}
\affiliation{%
  \institution{Tsinghua University}
  \country{}}

\author{Youyou Lu}
\affiliation{%
  \institution{Tsinghua University}
  \country{}}

\author{Jiwu Shu}
\authornote{Corresponding author.}
\affiliation{%
  \institution{Tsinghua University}
  \country{}}

\date{}

\begin{document}
\begin{abstract}
LLM agents are becoming a common interface for research, coding, and question
answering, yet their Thought--Action--Observation loop is often serial: the
model reasons, emits a tool call, then idles the GPU until the result returns.
This wait consumes 16--37\% of wall time in our workloads and 35--61\% in
prior reports.
Speculative tool execution can hide this wait, but existing systems need
auxiliary predictors, historical traces, or static workflow graphs, leaving a
gap for training-free, day-one deployment.
We observe that the model can be its own predictor: a probe forked at the
start of generation predicts Qwen3-32B's upcoming tool name with
74.6--99.6\% accuracy across five benchmarks.
We present \spork{} (\textbf{S}elf-s\textbf{P}eculative f\textbf{ORK}ing), a
training-free controller that dispatches the speculated tool call early,
overlapping its execution with the remaining chain-of-thought decode.
A cost model captures when speculation breaks even, and each component improves
one of its terms: a prefix-cache fork cuts probe cost, a confidence gate filters
mispredictions, and partial-token accept turns rejected probes into
speculative-decoding drafts.
On acceptance, the tool result is ready when reasoning ends; on rejection,
\spork{} falls back to serial execution with no correctness penalty.
On real-tool benchmarks, \spork{} cuts Qwen3-32B's GAIA $P_{95}$ by 18\%
(131.9$\to$108.1\,s); the mechanism holds across model sizes from 4B to 32B and
across dense and mixture-of-experts models, with task accuracy within 1\,pp of
baseline or better wherever measured.
\spork{} deploys as a thin controller over standard completion APIs (no
retraining, no auxiliary models, no offline traces) and is orthogonal to
token-level speculative decoding.
\spork{} is open source at \url{https://github.com/baihuajun24/spork}.
\end{abstract}

\maketitle
\thispagestyle{plain}

\section{Introduction}
\label{sec:intro}

\introepigraph
  {Treat things before they exist. Regulate things before disorder begins.}
  {Lao Tzu}
  {}

\noindent LLM-powered agents are becoming a common interface for complex work such as deep
research, coding, and tool-assisted question answering~\citep{IntroducingDeepResearch2025,
ManusHandsAI,ClaudeCodeClaude,GitHubCopilotYour2025,yao_react_2023,schick_toolformer_2024}.
In these systems, the LLM acts as the controller: it reasons over the user request,
emits a structured tool call, waits for the external tool to finish, and then resumes
generation using the tool result.
This \emph{Thought--Action--Observation} loop~\citep{yao_react_2023} is simple and safe,
but it is also often serial: each tool result conditions all subsequent tokens, so
the serving system cannot resume generation until the result arrives.
While the tool is running, the GPU often has no useful work to do for that agent turn.

This tool wait is hard to engineer away because it is structurally bounded by network
round-trips, container cold-starts, and database I/O that resist the exponential
improvements seen in model serving.
It is also large.
Our profiling across three benchmarks in Figure~\ref{fig:ttool_dist} shows that tool
execution accounts for 16--37\% of end-to-end wall time, and on turns with real network
tools the GPU is idle for the majority of the turn; \citet{sui_act_2026} report even
higher fractions, 35--61\% of total agent latency across agent workloads.

\begin{figure}[!t]
  \centering
  \includegraphics[width=0.92\columnwidth]{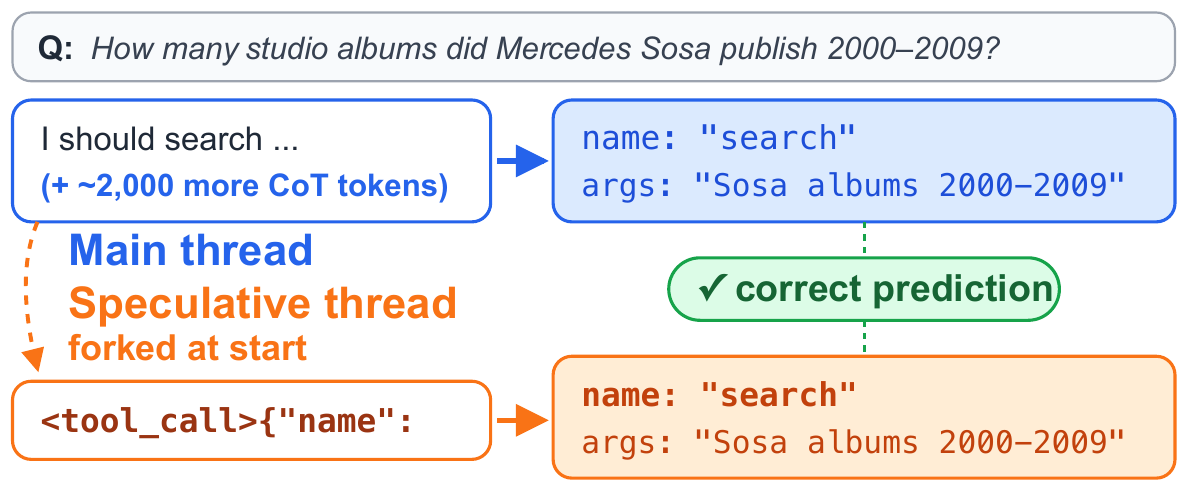}
  \caption{\textbf{Tool intent is visible early.}
  A speculative thread, forked at the main thread's first token with a forced tool-call
  prefix, predicts the exact tool call that the main thread emits ${\sim}2{,}000$ CoT
  tokens later.}
  \label{fig:teaser_prompt}
\end{figure}
\FloatBarrier

As LLM inference grows faster (through speculative decoding~\citep{leviathan_fast_2023},
hardware scaling, and disaggregated serving~\citep{zheng_sglang_2024}), the fraction of
wall time spent waiting for external tools \emph{increases}, not decreases, making
tool-wait the asymptotic bottleneck of agent latency.
Figure~\ref{fig:teaser_prompt} illustrates the opportunity this paper exploits: after a
single streaming token, a forked probe (sharing the same KV cache prefix) predicts the
identical web-search query that the main stream will emit after 2{,}000 tokens of
chain-of-thought.
The tool can be dispatched speculatively during those 2{,}000 tokens of decode, hiding
most of its network latency behind useful computation.

Existing systems attack adjacent parts of this problem but leave a gap for day-one,
\textit{training-free} deployment without predictors or historical traces.
Token-level speculative decoding~\citep{leviathan_fast_2023,li_eagle_2024,cai_medusa_2024}
and LLM serving systems~\citep{kwon_efficient_2023,zheng_sglang_2024,parrot24,
agrawal2024sarathiserve} reduce generation or serving overhead, but they do not dispatch
external tools earlier: the tool stall begins only \emph{after} the model has finished
emitting the call.
Workflow and serverless optimizers~\citep{orion,stojkovic2023specfaas,fu2024serverlessllm}
can prewarm or schedule known execution graphs, whereas tool-using agents generate their
next tool call online; there is no static graph to prefetch.

Recent agent-speculation systems move closer to execution-level overlap, but they rely on
a separate source of predictions: Speculative Actions~\citep{ye_speculative_2026} and
DualSpec~\citep{zhong_dualspec_2026} use predictors or verifier policies, while
PASTE~\citep{sui_act_2026} mines recurring tool-call patterns from historical traces.
These designs are powerful when traces, predictors, or repeated workflows are available;
they do not speculate from the currently running model alone.

\begin{figure}[!t]
  \centering
  \includegraphics[width=0.92\columnwidth]{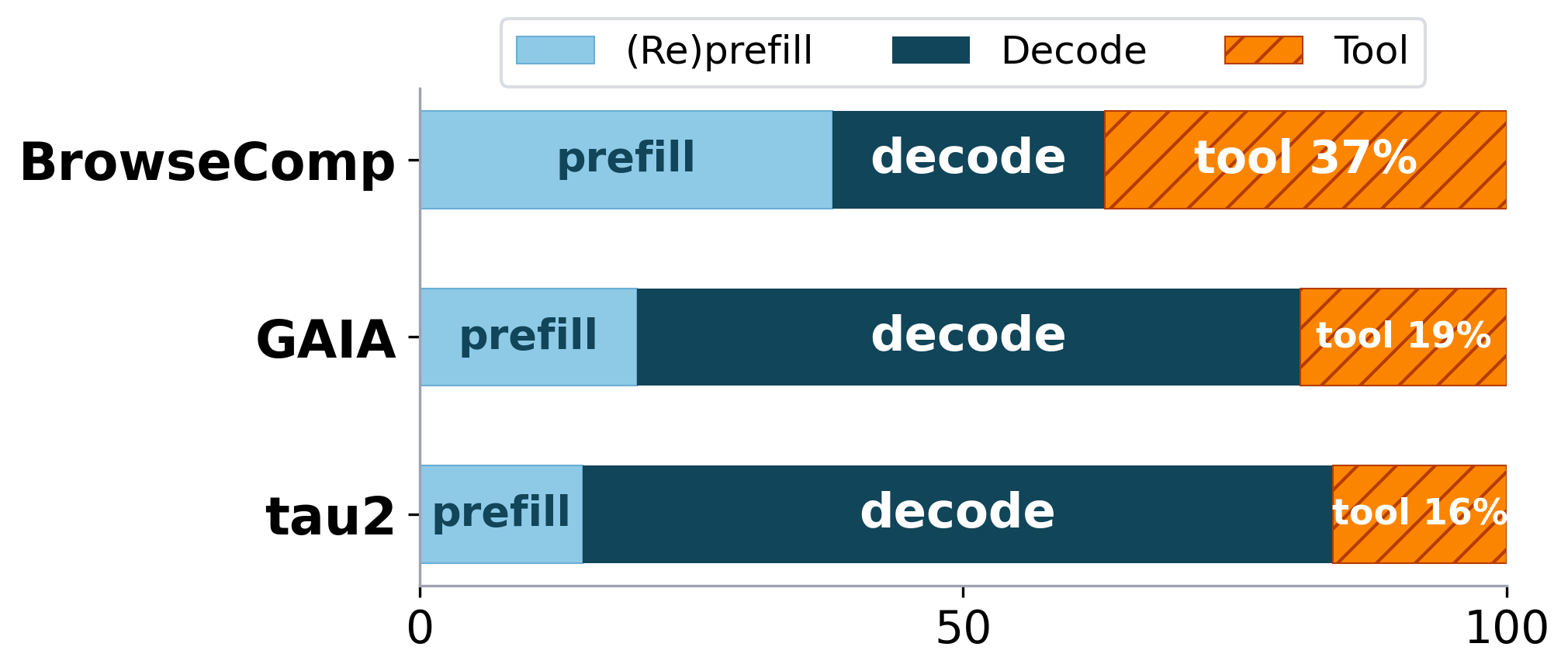}
  \caption{\textbf{Tool time is significant across agent tasks.} Across search, browsing, and database-style tasks, tool execution accounts for 16--37\% of end-to-end wall time.} %
  \label{fig:ttool_dist}
\end{figure}

The analogy to speculative execution in microprocessors is instructive: a CPU executes
past an unresolved branch and rolls back on misprediction.
The difference is the prediction source: where CPUs use branch predictors trained on
instruction history, \spork{} uses the running model itself, with no auxiliary model,
no trace collection, and no retraining.

\spork{} exploits a different opportunity: instruction-tuned models often reveal their
next tool call before completing the full chain-of-thought (\S\ref{sec:insight1}).
The fork also emits logprobabilities that indicate prediction confidence, and even failed
probes share a verified prefix with the final tool call.
These properties let the serving system turn the target model into its own speculative
tool predictor.

In this paper, we establish three empirical insights that make self-speculative tool use
viable.
First, Qwen3-32B~\citep{yang_qwen3_2025} predicts its own next tool call with high name accuracy (74.6--99.6\%
across GAIA~\citep{mialon_gaia_2024}, tau2-bench~\citep{barres_tau2_2025},
BrowseComp~\citep{wei_browsecomp_2025}, HotpotQA~\citep{yang_hotpotqa_2018}, and
BFCL~\citep{patil_bfcl_2025}) at the first-token fork point; the model already
``knows'' what it will do before finishing its reasoning (\S3.1).
Second, the probe's logprobability confidence separates correct from incorrect predictions
(0.90 vs.\ 0.46 mean span probability on GAIA), and argument accuracy climbs from 7.6\% to
97.5\% as more chain-of-thought context is observed (\S3.2).
Third, even when strict matching rejects a probe, the probe shares a median of 18 verified
tokens with the final tool call. Reused as a speculative-decoding \emph{draft}, these
tokens let the main stream skip re-decoding 35--50\% of the tool-call body (\S3.3).

\spork{} realizes this opportunity as a training-free controller for GPU-served tool
agents.
At each agent turn it \textbf{forks} the KV cache after the main generation's first token
(\textbf{D1}: near-zero overhead via prefix-cache sharing, reducing probe cost from 1.6\,s
to 0.35\,s), \textbf{gates} speculative dispatch on the probe's own logprob confidence
(\textbf{D2}: maximizes the accepted overlap by committing at the earliest confident
probe), and \textbf{recovers} verified prefix tokens from rejected probes via
speculative-decoding verification (\textbf{D3}: a median of 18 tokens recycled as drafts,
avoiding $\sim$0.6\,s of re-decode per rejected turn in offline analysis).
On a strict name-and-arguments match, \spork{} uses the pre-computed tool result;
otherwise it falls back to serial execution with no correctness penalty.

On GAIA (N=165, real web search, Qwen3-32B), the full \spork{} configuration (D1+D2+D3)
reduces median per-query latency by 10\% (P50: 34.7\,s $\to$ 31.2\,s) and tail latency by
18\% (P95: 131.9\,s $\to$ 108.1\,s) while preserving task accuracy (EM within 1\,pp of
baseline).
On HotpotQA (N=200, real Wikipedia API, Qwen3-32B), the full D1+D2+D3 configuration is
Pareto dominant: faster \emph{and} higher quality than the baseline simultaneously (P95
speedup 1.06$\times$, EM +1.5\,pp).
A latency sweep on tau2-bench (airline domain, N=43, simulated tool latency 0.5--5\,s) validates
the cost model: mean speedup grows monotonically with tool duration, from 1.09$\times$ at
a 0.5\,s tool floor to 1.18$\times$ at 5\,s, matching EQ1's prediction within 2\%.
The mechanism generalizes across model scale and architecture: it holds on Qwen3-4B (on
tau2; neutral elsewhere) and on
the Qwen3.5-35B-A3B~\citep{qwen_qwen35_2026} mixture-of-experts (16--20\% $P_{95}$ reduction across tau2, GAIA, and HotpotQA).
We also characterize when \spork{} does \emph{not} help: without thinking-mode
chain-of-thought the overlap window vanishes, and a model whose forced probe diverges
from its own tool-call format defeats self-speculation (\S6).

This paper makes the following contributions:
\begin{enumerate}
\item \textbf{Empirical insights (\S3).}
We show that Qwen3-32B predicts its own tool names with high accuracy (74.6--99.6\%) at the
fork point, exposes confidence through logprobabilities, and leaves
recoverable prefixes when strict matching rejects; we formalize the conditions under
which speculation breaks even via a cost model (EQ1, \S2.2).
\item \textbf{Training-free system design (\S4).}
We present three mechanisms (D1--D3) that realize speculative overlap within the standard
vLLM serving stack ($<$0.3\% TPOT overhead), requiring no model changes, no auxiliary
predictors, and no historical traces.
\item \textbf{End-to-end evaluation with real tools (\S6).}
On GAIA and HotpotQA with real network tools, \spork{} reduces $P_{95}$ latency by 6--18\%
while improving or preserving task accuracy, and the mechanism generalizes across
architecture (dense and MoE), with gains tapering to neutral at 4B scale on real-tool
benchmarks.
We validate the cost model against measured operating points and characterize when
speculation wins and when it does not.
\end{enumerate}

\section{Background \& Preliminaries}
\label{sec:background}

\subsection{The Serial Tool-Wait Bottleneck}
\label{sec:bottleneck}

The dominant execution model for LLM agents follows the ReAct
loop~\citep{yao_react_2023}: the model reasons (Thought), emits a structured tool
invocation (Action), waits for the external service to return a result (Observation),
and then resumes reasoning conditioned on that result.
This Thought--Action--Observation cycle enforces a strict serial dependency:
\[
[\text{(Re)prefill}] \;\to\; [\text{Decode}] \;\to\; [\text{Tool wait}] \;\to\; \cdots
\]
Each turn (re)prefills the context (the user prompt on the first turn, the appended tool
result thereafter), decodes the reasoning and tool call, then stalls on the tool wait.
The tool call cannot be dispatched until decoding completes; the next decode cannot begin
until the tool result arrives and is (re)prefilled into the context window.
This creates a structural pipeline stall regardless of how fast individual components are.

\parabf{Tool latency dominates agent wall time.}
Figure~\ref{fig:ttool_dist} shows that tool execution consumes a substantial share of
end-to-end agent latency across three workloads: 16\% on tau2-bench (enterprise database
APIs with 2\,s simulated floor), 19\% on GAIA (real web search APIs), and 37\% on
BrowseComp (web search + page browse).
On BrowseComp, the distribution is heavily right-skewed: median tool latency is 1.19\,s
but $P_{95}$ reaches 70\,s, meaning tail-latency optimization requires addressing tool
stalls, not just model serving.

\parabf{Why existing solutions fall short.}
Token-level speculative decoding~\citep{leviathan_fast_2023,li_eagle_2024,cai_medusa_2024}
reduces generation latency but does not dispatch external tools earlier.
Parallel tool calling (e.g., OpenAI function calls) helps only when the model can identify
independent tools upfront; empirically, only 15--25\% of tool calls in GAIA and SWE-bench~\citep{jimenez_swebench_2024}
are parallelizable with their predecessor.
Workflow optimizers~\citep{orion,stojkovic2023specfaas,fu2024serverlessllm} prewarm known
execution graphs, but agent tool calls are generated \emph{online}: there is no static
graph to prefetch.
Recent agent-speculation systems~\citep{ye_speculative_2026,zhong_dualspec_2026,sui_act_2026}
move closer to execution-level overlap but require separate predictors, verifier policies,
or historical traces.
None speculates from the currently running model alone.

\subsection{Cost Model for Speculative Overlap}
\label{sec:eq1}

If we could predict the next tool call before the model finishes generating it, we could
dispatch the tool speculatively and overlap its execution with ongoing model decode.
The realized speedup depends on four measurable parameters:
\begin{itemize}
  \item $\alpha$: the fraction of tool-call \emph{turns} where speculation is accepted
        (gate rate). $\alpha$ is always measured per turn, not per dispatched probe; a
        retrying gate may issue several probes within one turn.
  \item $\toverlap$: mean realized time overlap per accepted turn (how much tool latency
        is hidden behind decode).
  \item $\Tbasestar$: the realized main-stream decode cost on the speculative path. It
        equals the baseline $\Tbase$ unless prefix recovery is active, which lowers it by
        letting the main stream skip re-decoding the verified tool-call tokens it recovers
        from the rejected probe on a miss.
  \item $\Toh$: per-turn overhead from probe computation and wasted speculative execution
        on rejected turns.
\end{itemize}

\noindent The resulting latency-ratio formula (full derivation in Appendix~\ref{app:eq1}):
\begin{equation}
  \boxed{\mathrm{Ratio} = \frac{\Tbase}{\Tbasestar - \alpha\cdot\toverlap + \Toh}}
  \label{eq:eq1}
\end{equation}

\noindent Speculation improves latency when $\mathrm{Ratio} \geq 1$, which approximately
requires $\alpha\cdot\toverlap \geq \Toh$ (exact form in Appendix~\ref{app:eq1}): the time
saved by overlapping tool execution on accepted turns must exceed the overhead incurred
across all turns.
The upper bound is set by the tool-time fraction $\ftool$: with perfect prediction
($\alpha=1$, $\Toh=0$), $\mathrm{Ratio} = 1/(1-\ftool)$.
For BrowseComp ($\ftool=0.366$), this ceiling is $1.58\times$.

The next section establishes empirically that these parameters are achievable: $\alpha$
is high, $\Toh$ is low, and even rejected probes contribute via prefix recovery.

\section{Self-Speculative Tool Use}
\label{sec:selfspec}

The cost model in \S\ref{sec:eq1} identifies three conditions for speculation to succeed:
high gate acceptance ($\alpha$), sufficient overlap time ($\toverlap$), and low overhead
($\Toh$).
Whether these are achievable depends on an empirical question about the model itself:

\begin{tcolorbox}[
  colback=blue!3!white,
  colframe=blue!45!black,
  boxrule=0.45pt,
  arc=1.5pt,
  left=4pt,
  right=4pt,
  top=3pt,
  bottom=3pt
]
\textbf{Research question.}
Can a running LLM predict its own next tool call \emph{accurately} enough,
\emph{early} enough, and \emph{recoverably} enough that speculative tool use
pays off?
\end{tcolorbox}

\noindent We constrain the prediction source to the running model's own
state: no auxiliary model, no historical traces.
We answer affirmatively with three empirical insights, established on
\textbf{Qwen3-32B} across five real-tool and simulated benchmarks (GAIA, HotpotQA,
tau2-bench, BrowseComp, BFCL): the model already knows its next tool call
(Insight~1, \emph{accuracy}), its own confidence signals when to commit
(Insight~2, \emph{timing}), and even rejected probes leave verified prefixes
(Insight~3, \emph{recovery}). How these properties scale to a smaller model (Qwen3-4B)
and where they break (native XML tool formats) is examined in \S\ref{sec:envelope}.

\subsection{Insight 1: Models Already Know Their Next Tool Call}
\label{sec:insight1}

A short forced-prefix probe can expose the next tool identity well before the serial tool
boundary.
We force a tool-call prefix immediately after the model's first decode token
(fork-at-start) and compare the probe's predicted tool name and arguments against the
model's unrestricted generation.

\begin{figure}[t]
  \centering
  \includegraphics[width=0.95\columnwidth]{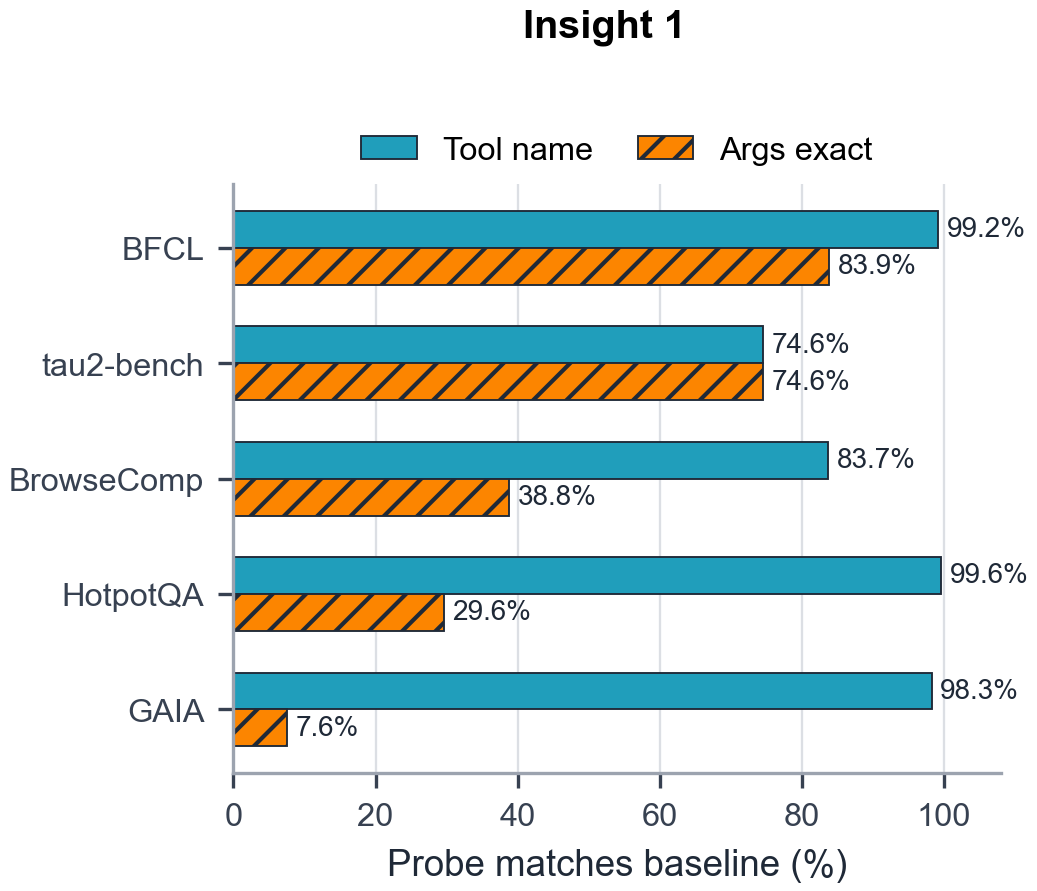}
  \caption{\textbf{Fork-at-start probes recover the next tool name across workloads.}
  On Qwen3-32B, forced tool-call probes near decode start predict the eventual tool
  name across five benchmarks (sorted by argument exact-match; hatched bars).
  Protocols: replay pos=0 (GAIA, tau2-bench, BFCL) and online first-token
  (BrowseComp, HotpotQA).}
  \label{fig:fig1b_fork_acc}
\end{figure}

\parabf{Foresight across benchmarks.}
On Qwen3-32B, near-decode-start tool-name accuracy is consistently high across the five
benchmarks in Figure~\ref{fig:fig1b_fork_acc}: 98.3\% on GAIA, 74.6\% on tau2-bench,
83.7\% on BrowseComp, 99.6\% on HotpotQA, and 99.2\% on BFCL.
Argument-exact accuracy is lower and more variable (7.6--83.9\%), reflecting the
difficulty of predicting precise parameter values before observing the chain-of-thought.
This is the gap that Insight~2 addresses.

\parabf{No-op recovery.}
On 11 tau2-bench turns where the baseline declined to call any tool, 9/11 forced-probe
outputs were correct tool calls the model should have made; fork-prefix can act as a
cheap correction against model over-caution.

\parabf{Implication for EQ1.}
High name accuracy means $\alpha$ (gate acceptance on name match) is intrinsically high
before any confidence gating.
The remaining challenge is argument accuracy, which Insight~2 addresses.

\subsection{Insight 2: Confidence Improves with Chain-of-Thought}
\label{sec:insight2}

\begin{figure}[!t]
  \centering
  \includegraphics[width=0.93\columnwidth]{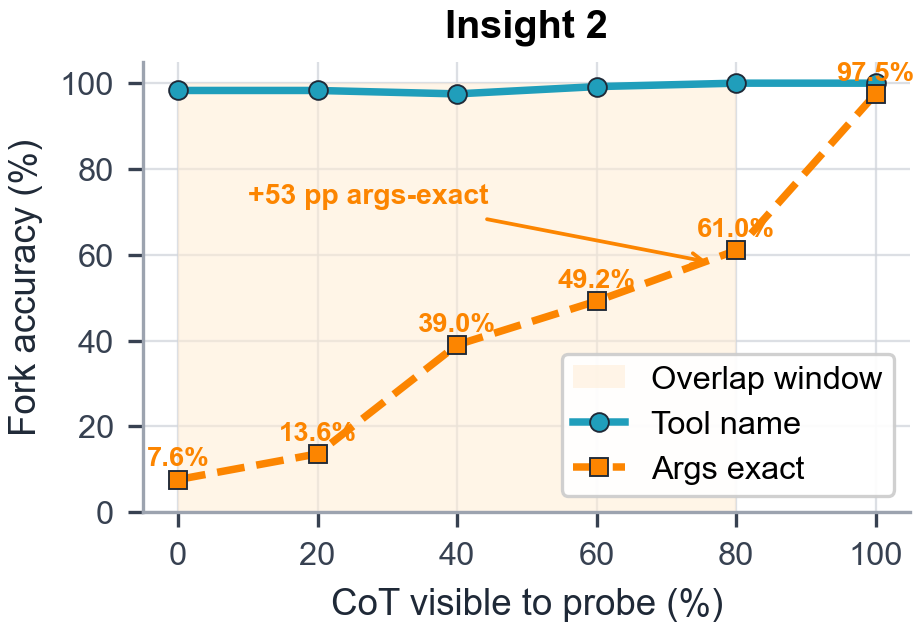}
  \caption{\textbf{Fork accuracy rises monotonically with CoT context.}
  On GAIA (Qwen3-32B), args-exact accuracy climbs from 7.6\% at fork-at-start to
  61.0\% at 80\% CoT and 97.5\% at think-end.
  Later probes are more accurate but leave a smaller overlap budget.}
  \label{fig:cot_curve}
\end{figure}

The useful operating point is not necessarily the latest or most accurate probe; it is
the earliest probe whose confidence justifies committing tool execution time.

\parabf{Logprob separability.}
The probe's own generation logprobabilities over the tool-name span separate correct
from incorrect predictions without any classifier training.
On GAIA (Qwen3-32B, think mode): the mean span probability is 0.90 for correct
predictions versus 0.46 for incorrect ones.
This makes confidence observable without external classifiers: low confidence probes can
be delayed or rejected before they spend real tool time.
On tau2-bench no-think mode the signal saturates (correct $\approx$ incorrect $\approx$
0.979): tool selection is a first-token reflex, so confidence is most informative on
reasoning-mode workloads with diverse tool sets.

\parabf{CoT length amplifies accuracy.}
Fork accuracy at six positions along the CoT trajectory (0--100\% of tokens, gold-CoT
replay) rises monotonically (Figure~\ref{fig:cot_curve}).
On GAIA (N=118 tool-call questions, Qwen3-32B), argument-exact accuracy climbs from
7.6\% at position~0 to 61.0\% at 80\% CoT and 97.5\% at think-end, a
\textbf{+53\,pp gap} that a later probe can exploit, at the cost of a smaller overlap
budget (the decode time remaining to hide the tool, distinct from EQ1's realized
$\toverlap$).

\parabf{Implication for EQ1.}
The accuracy--timeliness tradeoff means a \emph{dynamic} gate is better than a fixed fork
position: commit early when confident, wait when uncertain.
This is the foundation of D2's adaptive confidence gate (\S\ref{sec:d2}).

\subsection{Insight 3: Rejected Probes Still Contain Verified Prefixes}
\label{sec:insight3}

Strict matching is necessary for correctness, but rejection does not mean the probe was
useless: a rejected probe can serve as a \emph{draft} that the main stream verifies and
partially reuses via speculative decoding.

\begin{figure}[t]
  \centering
  \includegraphics[width=0.93\columnwidth]{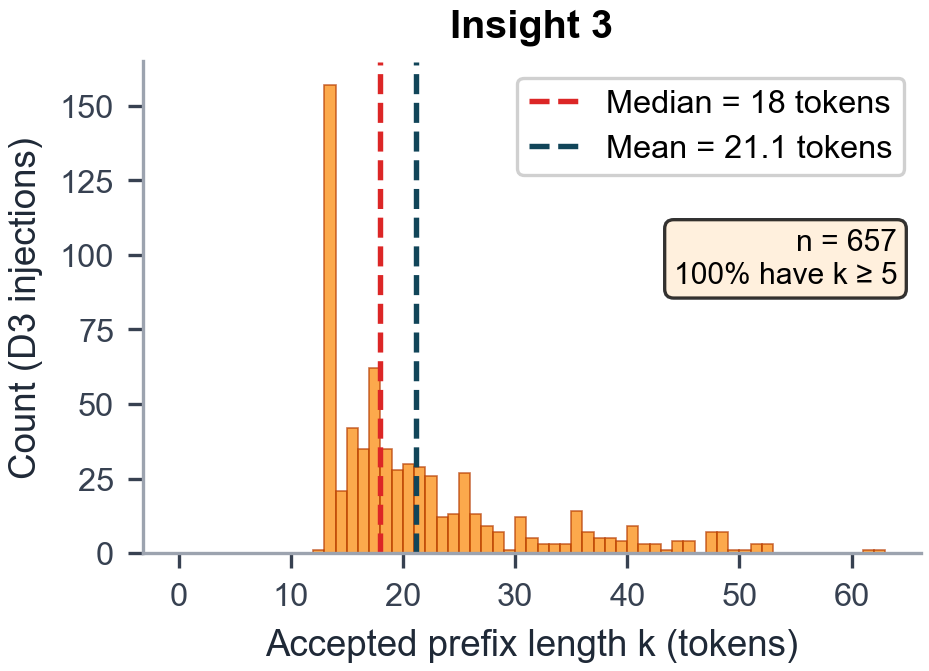}
  \caption{\textbf{Most rejected probes share recoverable prefixes.}
  On GAIA, the 25th-percentile shared prefix is 14 tokens: 75\% of rejected probes
  share at least 14 tokens with the main tool call.}
  \label{fig:insight3_prefix}
\end{figure}

\parabf{Prefix overlap after strict rejection.}
When the strict gate rejects a probe (tool name or arguments differ from main), the
probe's tool-call token sequence still shares a long verified prefix with the eventual
main output.
Figure~\ref{fig:insight3_prefix} quantifies this on two GAIA N=165 seeds
($n=657$ rejected-probe events): the median shared prefix is \textbf{18 tokens},
with a mean of 21.1 and 25th/75th percentiles of 14 and 25 tokens (often 35--50\% of
the full tool-call body).
A rejected probe need not be pure waste: if the shared prefix can be verified, the main
generation can reuse those tokens and decode only the remaining suffix.

\parabf{Implication for EQ1.}
Prefix recovery reduces the effective $\Toh$ on the miss branch: instead of discarding
the probe and re-decoding the full tool call from scratch (${\sim}50$ tokens), the system
verifies and accepts $k$ matching tokens and decodes only the $50-k$ suffix.
This is the foundation of D3 (\S\ref{sec:d3}).

\medskip
\begin{tcolorbox}[
  colback=blue!3!white,
  colframe=blue!45!black,
  boxrule=0.45pt,
  arc=1.5pt,
  left=4pt,
  right=4pt,
  top=3pt,
  bottom=3pt
]
\textbf{Summary.}
The three insights establish that self-speculative tool use is \emph{feasible},
\emph{steerable}, and \emph{recoverable}: the model predicts its next tool identity
early, its own confidence signals when to commit, and even rejected probes yield
verified prefixes.
\end{tcolorbox}

\noindent The next section translates these insights into three system mechanisms that
jointly push the operating point toward the EQ1 break-even condition from both sides.

\section{\spork{} Design}
\label{sec:design}

\begin{tcolorbox}[
  colback=blue!3!white,
  colframe=blue!45!black,
  boxrule=0.45pt,
  arc=1.5pt,
  left=4pt,
  right=4pt,
  top=3pt,
  bottom=3pt
]
\textbf{Systems question.}
How can we overlap tool execution with main generation at minimal added overhead,
while keeping the agent lossless with respect to its serial execution?
\end{tcolorbox}

\noindent \spork{} answers with three mechanisms: \textbf{Design~1 (D1)},
\textbf{Design~2 (D2)}, and \textbf{Design~3 (D3)}, each grounded in one insight
from \S\ref{sec:selfspec} and targeting a specific parameter of the cost model
(EQ1, \S\ref{sec:eq1}).

\begin{figure*}[!t]
  \centering
  \includegraphics[width=\textwidth]{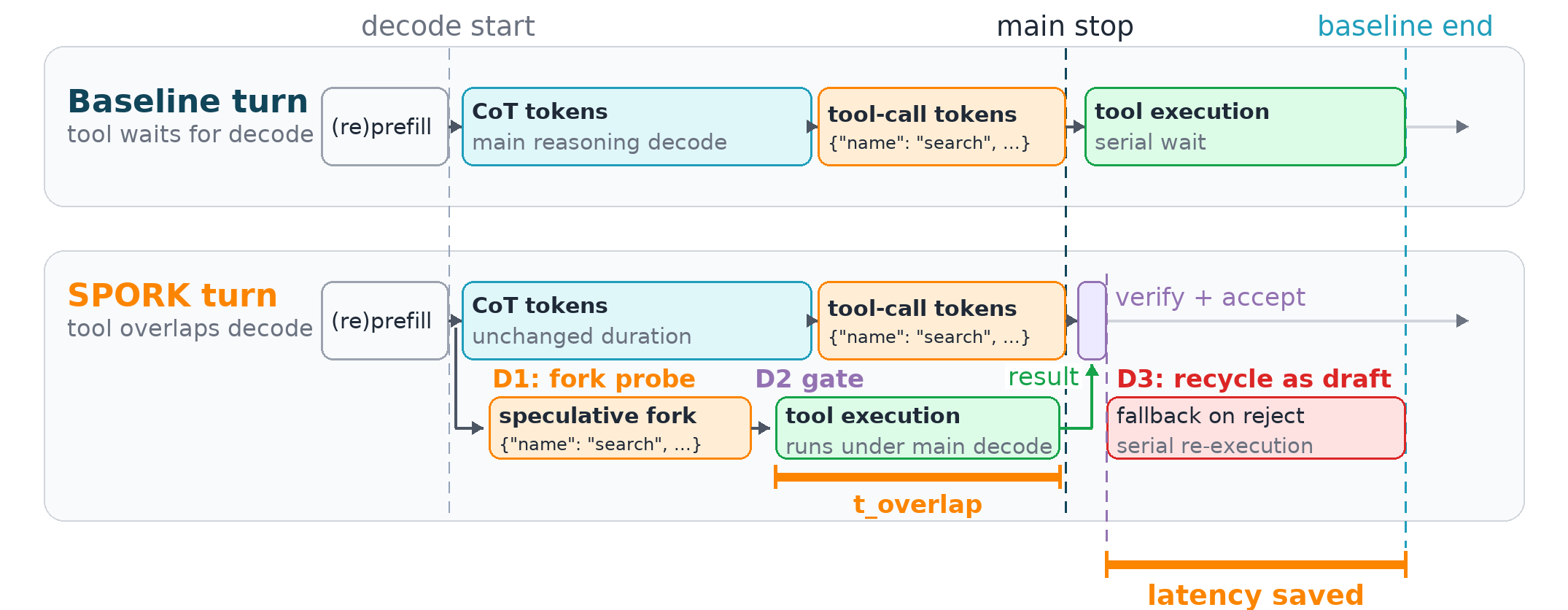}
  \caption{\textbf{\spork{} method overview.} \textit{Top:} the baseline turn is strictly
  serial: (re)prefill, CoT decode, tool-call decode, then a stall while the tool
  executes. \textit{Bottom:} \spork{} forks the KV cache after the main's first token and
  launches a forced tool-call probe (\textbf{D1}); once the probe's logprob confidence
  clears the gate (\textbf{D2}), the predicted tool executes during the remaining decode,
  producing the overlap $t_{\mathrm{overlap}}$. A strict match at main-stream completion
  accepts the pre-computed result; on mismatch the turn falls back to serial execution and
  \textbf{D3} recycles the rejected probe's tokens as a speculative-decoding draft.}
  \label{fig:method_overview}
\end{figure*}

At each agent turn, \spork{} runs the following pipeline (Figure~\ref{fig:method_overview}):
\begin{enumerate}
\item Dispatch the main generation as a streaming chat-completion request.
\item After the first streaming token, launch a fork thread that issues a raw
      completion request with a forced tool-call prefix.
\item The fork thread monitors its generation logprobs (D2); once confidence reaches a
      threshold $\theta$ (the confidence gate, \S\ref{sec:d2}), dispatch speculative
      tool execution.
\item When the main stream reaches its tool call, compare it against the committed
      probe under the gate criterion. On match: use the pre-computed result.
      On mismatch: fall back to serial tool execution. In the engine configuration,
      D3 has meanwhile served the probe's tool-call body to the main stream as
      speculative-decoding draft tokens, so the main decode re-emits only the tokens
      after the first mismatch rather than the whole call.
\end{enumerate}

The D1/D2 controller is a Python application-layer component that uses standard serving
APIs and collects no training data.
D3 additionally integrates with vLLM's speculative-decoding proposer path to inject
verified probe tokens as drafts; we describe this as a prototype integration rather than
a generic no-engine-modification claim.

\parabf{EQ1 parameter mapping.}
Each mechanism targets a different term in the cost model
(Table~\ref{tab:design_rationale}; Eq.~\ref{eq:eq1}, \S\ref{sec:eq1}):
D1 follows Insight~1 and \emph{minimizes the overhead term} $\Toh$: by reusing the
main's prefix KV cache, the probe's cost collapses from 1.6\,s to 0.35\,s, and its
dispatch at the first streaming token opens the overlap window.
D2 follows Insight~2 and \emph{maximizes the expected overlap} $\alpha\cdot\toverlap$:
gating dispatch on the probe's own logprob confidence, it commits at the earliest probe
confident enough to act on, trading a slightly smaller window for much higher
acceptance.
D3 follows Insight~3 and lowers $\Tbasestar$ on misses by recycling the probe's verified
prefix as draft tokens, so the main stream re-decodes only the suffix after the first
mismatch.
Together, all three push the operating point toward the $\alpha\cdot\toverlap \geq \Toh$
break-even condition from both sides.

\begin{table}[tbp]
\caption{\textbf{Design rationale.} Each mechanism is tied to one insight and one
EQ1 term.}
\label{tab:design_rationale}
\centering
\small
\begin{tabular}{@{}lll@{}}
\toprule
Design & Insight & EQ1 effect \\
\midrule
\textbf{D1} & Insight~1 & minimizes $\Toh$ \\
\textbf{D2} & Insight~2 & maximizes $\alpha\cdot\toverlap$ \\
\textbf{D3} & Insight~3 & lowers $\Tbasestar$ (miss re-decode) \\
\bottomrule
\end{tabular}
\end{table}

\spork{} composes the three mechanisms. D1+D2 form the training-free controller, and D3 adds
engine-side partial-token recovery. The ablation in \S\ref{sec:ablation} measures each
mechanism's contribution.
The controller is stateless across turns except for the shared prefix cache inside the
serving engine: each agent turn forks independently, and the gate always compares against
the main stream's final tool call before any result is committed to the conversation history.

\subsection{D1: Prefix-Cache Fork for Near-Zero Probe Overhead}
\label{sec:d1}

\begin{figure}[t]
  \centering
  \includegraphics[width=0.95\columnwidth]{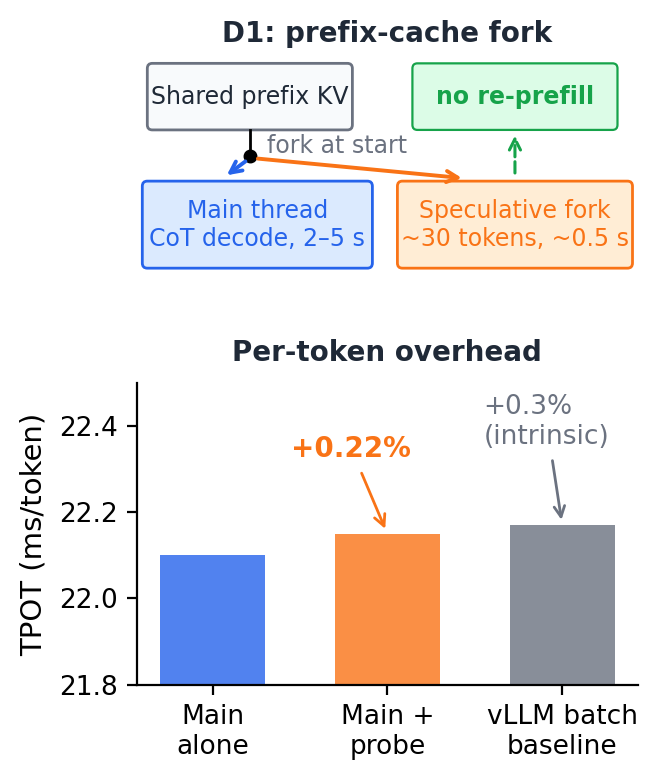}
  \caption{\textbf{D1 reuses the prefix KV cache to keep probe overhead small.}
  Probe TPOT overhead (+0.22\%) is within intrinsic batching noise (+0.3\%).}
  \label{fig:d1_overhead}
\end{figure}

\parabf{Mechanism.}
The probe is not free, since it must generate a tool call of its own; its overhead
decomposes into a decoding cost and a prefilling cost.
The decoding cost is small.
The fork decodes only the tool-call body (typically $\leq 50$ tokens, stopping at the
tool-call close), $10\text{--}50\times$ shorter than the main's CoT, about 0.3\,s on
Qwen3-32B at 15K context.
The prefilling cost is the heavy part: the fork shares the full prompt prefix
(system + user + history) with the main request, and re-prefilling it costs
$\sim$1.3\,s at the same context; worse, we measure that dispatching the fork
simultaneously with the main request causes 2.5$\times$ prefill contention.
Modern LLM serving systems such as vLLM~\citep{kwon_efficient_2023} and
SGLang~\citep{zheng_sglang_2024} cache prefix KV states; D1 therefore
dispatches the fork \emph{after} the main's first streaming token.
Because the main request has by then populated the prefix cache, the fork's prefill is
a cache hit at near-zero cost ($\sim$0.05\,s), and the staggered dispatch reduces
prefill interference to $<1\%$.
With D1, probe overhead collapses from 1.6\,s (1.3\,s prefill + 0.3\,s decode) to
0.35\,s (0.05\,s cache-hit prefill + 0.3\,s decode).
Figure~\ref{fig:d1_overhead} shows the resulting end-to-end effect: the fork's measured
TPOT overhead on the main stream is within intrinsic batching noise.

\subsection{D2: Adaptive Confidence Gate}
\label{sec:d2}

\parabf{Motivation.}
Without gating, D1 dispatches speculative tool execution on \emph{every} fork, regardless
of whether the probe is likely correct.
Each wasted dispatch consumes network/compute resources and, on side-effecting tools, must
be rolled back.
Insight~2 (\S\ref{sec:selfspec}) shows that probe accuracy improves dramatically with
context: argument accuracy rises from 7.6\% at 0\% CoT to 97.5\% at 100\% CoT
(Figure~\ref{fig:cot_curve}).
D2 exploits this by \emph{not committing} until the fork's own token-level confidence
signals that the generated tool call is likely correct, enabling the system to filter
out low-quality probes before they trigger wasted work.

\parabf{Confidence metric.}
The fork generates tool-call tokens with per-token logprobs enabled.
We define the confidence score as the \emph{minimum top-1 token probability} over the
tool-name span (the first $L$ tokens after the \texttt{"name":~"} prefix):
\begin{equation}
  c = \min_{i=2}^{L}\; \exp(\ell_i),
  \label{eq:confidence}
\end{equation}
where $\ell_i$ is the log-probability of the top-1 token at position $i$, and $L$ is
the length of the generated tool name (typically 1--4 tokens).
The minimum starts at $i=2$ because the first token of the span (the opening quote)
is always high-probability boilerplate.
The minimum operator is deliberately conservative: a single low-confidence token in the
name span signals that the model is uncertain about the tool selection, and committing
would likely waste a tool execution.

\parabf{Gate logic.}
Once $c \geq \theta$, the fork commits to speculative tool execution.
If the fork finishes without $c$ reaching $\theta$, no speculative dispatch occurs and
the turn falls back to serial execution at zero additional cost (the fork decoding is
already overlapped with main CoT).

\parabf{Strict acceptance.}
\spork{} accepts the probe if and only if its full tool call (name \emph{and} serialized
arguments) matches the main generation exactly.
We deliberately do not ship a name-only gate: a name-only match can silently change tool
behavior on side-effecting tools, and D3 (\S\ref{sec:d3}) already recovers most of the
savings that name-only acceptance would deliver without compromising correctness.

\begin{figure}[t]
  \centering
  \includegraphics[width=0.86\columnwidth]{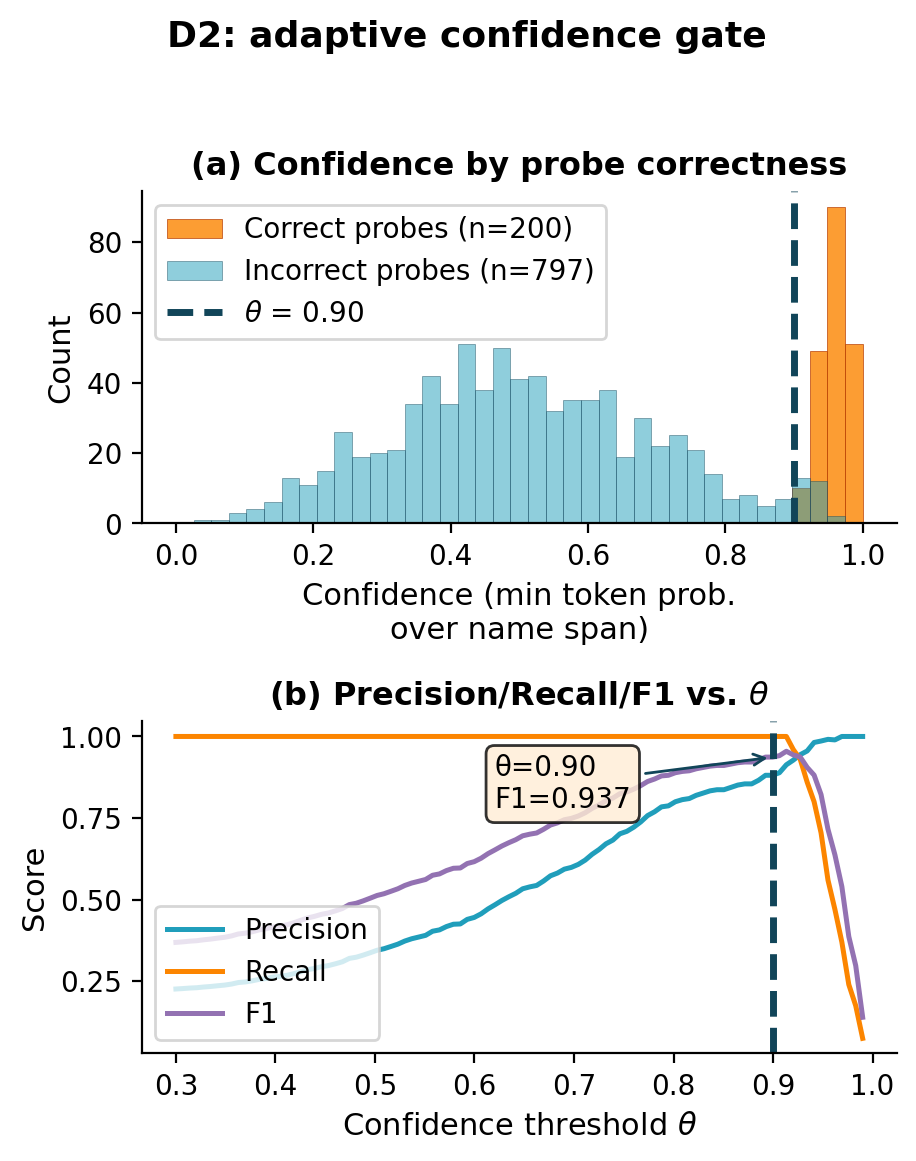}
  \caption{\textbf{D2 confidence gate threshold selection} (GAIA N=127, 997 probes).
  \textit{(a)}~Correct probes cluster above $\theta{=}0.90$; incorrect probes spread
  below. The threshold cleanly separates them.
  \textit{(b)}~Precision/Recall/F1 vs.\ $\theta$: the operating point $\theta{=}0.90$
  maximizes F1 (0.937) with 88\% precision and 100\% recall.
  D2 filters 77\% of probes (saving wasted tool executions) while committing every
  probe that would have been correct.}
  \label{fig:d2_gate}
\end{figure}

\parabf{Threshold selection.}
Figure~\ref{fig:d2_gate} shows the confidence distribution across 997 probes on GAIA
(N=127, Qwen3-32B).
Correct probes (those whose tool name matches the main generation) cluster tightly above
$c=0.9$, while incorrect probes spread broadly over $[0.2, 0.9]$.
At $\theta=0.90$, the gate achieves 88\% precision with 100\% recall
(F1\,$=$\,0.937), filtering 77\% of all probes while retaining every correct one.
We select $\theta=0.90$ as the operating point that maximizes F1.

\parabf{Alternatives considered.}
(1)~\emph{Fixed token-count delay} (wait $N$ main tokens before forking): rigid, does not
adapt to per-task difficulty, and delays correct probes unnecessarily.
(2)~\emph{Entropy-based gate} (threshold on next-token entropy): entropy over the full
vocabulary correlates poorly with tool-name correctness; min-prob over the specific name
span is a more direct signal.
(3)~\emph{Learned classifier}: adds inference latency and a training dependency; the
zero-parameter min-prob gate achieves F1\,$>$\,0.93 with no training.

\parabf{Effect on EQ1.}
D2 maximizes the expected overlap $\alpha\cdot\toverlap$: by waiting for the earliest
confident probe it trades a slightly smaller window for much higher acceptance, and by
filtering low-confidence turns before dispatching \texttt{spec\_tool} it avoids wasted
work (on BrowseComp, 52\% of strict-gate turns are rejected and never trigger a tool).
Under the strict gate without D2, $\alpha \approx 0.22$ on GAIA;
with D2, $\alpha$ rises to $0.37$ (49 vs.\ 101 accepted probes).
Recall that $\alpha$ counts tool-call \emph{turns} (\S\ref{sec:eq1}); the dispatched-probe
totals are larger (434 for D1, 2124 for D1+D2) because the retry cadence can issue several
probes within one turn.

\subsection{D3: Partial-Token Accept for Rejected Probes}
\label{sec:d3}

When the strict gate rejects a probe, the main thread must otherwise generate the full tool
call from scratch (${\sim}50$ tokens, ${\sim}1.5$\,s at 32B scale).
Empirical analysis of 1268 probe--baseline pairs (BrowseComp,
Appendix~\ref{app:d3}) shows that even rejected probes match
the main's greedy output for the first $k$ tokens: mean $k \approx 29.5$, median 27,
$P_{25}=17$, $P_{75}=35$.

\begin{figure}[t]
  \centering
  \includegraphics[width=0.95\columnwidth]{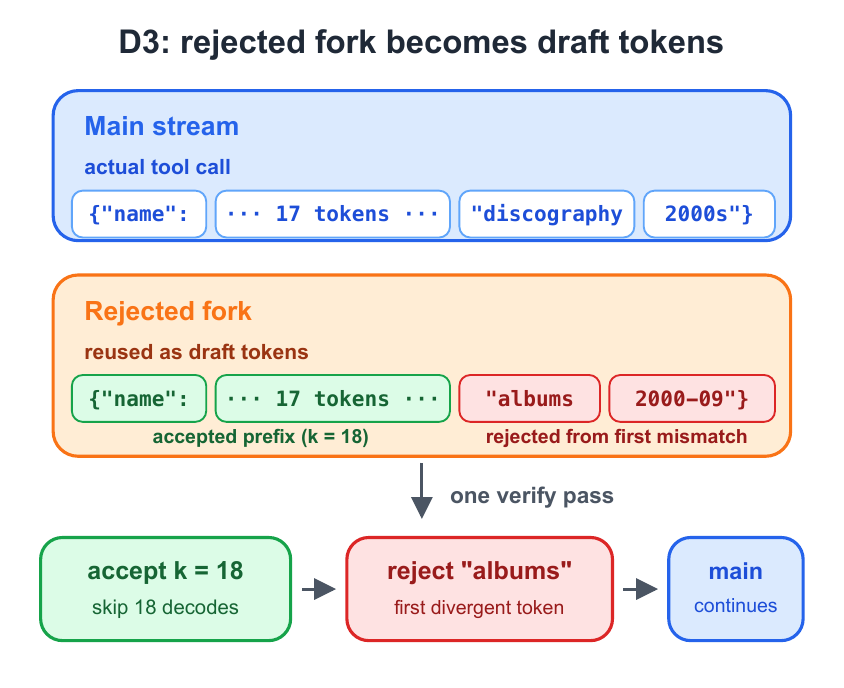}
  \caption{\textbf{D3 treats the rejected fork as a draft for speculative verification.}
  Instead of discarding the probe, \spork{} verifies its tool-call tokens, accepts the
  longest matching prefix, and lets the main stream decode only the suffix after the first
  mismatch.}
  \label{fig:d3_spec_dec}
\end{figure}

\parabf{Mechanism.}
D3 turns the probe into draft tokens for the main stream's \emph{own} speculative
decoding rather than re-decoding the tool call from scratch
(Figure~\ref{fig:d3_spec_dec}). We integrate a
\texttt{SporkProposer} into the serving engine's speculative-decoding proposer path:
as the main request decodes, the proposer watches for the \texttt{<tool\_call>}
boundary and, at that point, supplies the probe's tool-call body as the draft for
that request.
\begin{enumerate}
  \item When the main stream reaches the tool-call opener, the probe body is
        registered as the per-request draft.
  \item The engine verifies the draft against the main model in the same decode step
        (no separate verification pass, no second prompt); it accepts the longest
        prefix matching the main model's greedy tokens.
  \item The first $k$ matching tokens are accepted; the main stream decodes only the
        tool-call tokens after the first mismatch.
\end{enumerate}
D3 runs only in the in-process (engine) configuration, whose baseline is ngram
speculative decoding; the HTTP controller configuration (\S\ref{sec:implementation})
runs D1+D2 only.
D3 reduces the per-rejected-turn cost to approximately
$\Tbasestar + \Toh$, where $\Tbasestar = \Tbase - T_{\mathrm{D3}}$ is the realized base
decode after recycling the verified prefix as drafts ($\approx$0.6\,s per rejected turn,
$\approx$0.31\,s amortized per turn on BrowseComp; Appendix~\ref{app:d3}).
D3 is complementary to D1+D2: tool-level speculation on accepted turns, token-level
recycling on rejected turns; the fork itself is the draft, avoiding a separate draft model.

\section{Implementation}
\label{sec:implementation}

\parabf{Overview.}
\spork{} is implemented in Python as a controller layer plus an optional D3 proposer
integration.
The D1/D2 path (fork dispatch, logprob monitoring, asynchronous tool execution, and
gate verification) uses standard OpenAI-compatible vLLM endpoints
(\texttt{/v1/chat/completions} for the main stream, \texttt{/v1/completions} for the fork).
D3 adds a small \texttt{SporkProposer} integration with vLLM's speculative-decoding
proposer path so that verified probe tokens can be injected as draft tokens at the
tool-call boundary.
Algorithm~\ref{alg:spork} summarizes the end-to-end control flow.
The HTTP controller realizes D1+D2; the engine configuration additionally performs D3,
which the HTTP path omits.

\begin{algorithm}[t]
\caption{\spork{} Controller with D1+D2+D3}
\label{alg:spork}
\begin{algorithmic}[1]
\REQUIRE User request, tool set $\mathcal{T}$, threshold $\theta$, retry budget $R$, token step $s$
\STATE Dispatch main stream (streaming chat completions)
\STATE Wait for main's first streaming token
\STATE $\mathrm{spec} \gets \emptyset$;\ $(\hat{t},\hat{a}) \gets \emptyset$
\FOR{$r = 0,1,\dots,R-1$}
\STATE Wait until main has decoded $r\cdot s$ further CoT tokens
\STATE Fork a completions request with forced prefix over observed CoT \COMMENT{D1 cache hit}
\STATE $(\hat{t},\hat{a}) \gets \mathrm{parse}(\mathrm{fork})$;\ $c \gets \min_{j \in \mathrm{name\_span}} \exp(\ell_j)$ \COMMENT{D2}
\IF{$(\hat{t},\hat{a})$ valid \AND $c \geq \theta$}
\STATE $\mathrm{spec} \gets \mathrm{async\_tool}(\hat{t},\hat{a})$ \COMMENT{commit; overlaps main decode}
\STATE \textbf{break}
\ENDIF
\IF{main stream finished}
\STATE \textbf{break}
\ENDIF
\ENDFOR
\STATE Wait for main tool call $(t^*,a^*)$
\IF{$\mathrm{spec}\neq\emptyset$ \AND $(\hat{t},\hat{a}) = (t^*,a^*)$}
\STATE \textbf{return} pre-computed $\mathrm{spec}$ result \COMMENT{strict match}
\ELSE
\STATE \textbf{return} $\mathrm{tool}(t^*,a^*)$ \COMMENT{serial fallback; D3 draft-injects the probe body during decode}
\ENDIF
\end{algorithmic}
\end{algorithm}

\begin{figure*}[t]
  \centering
  \includegraphics[width=0.84\textwidth]{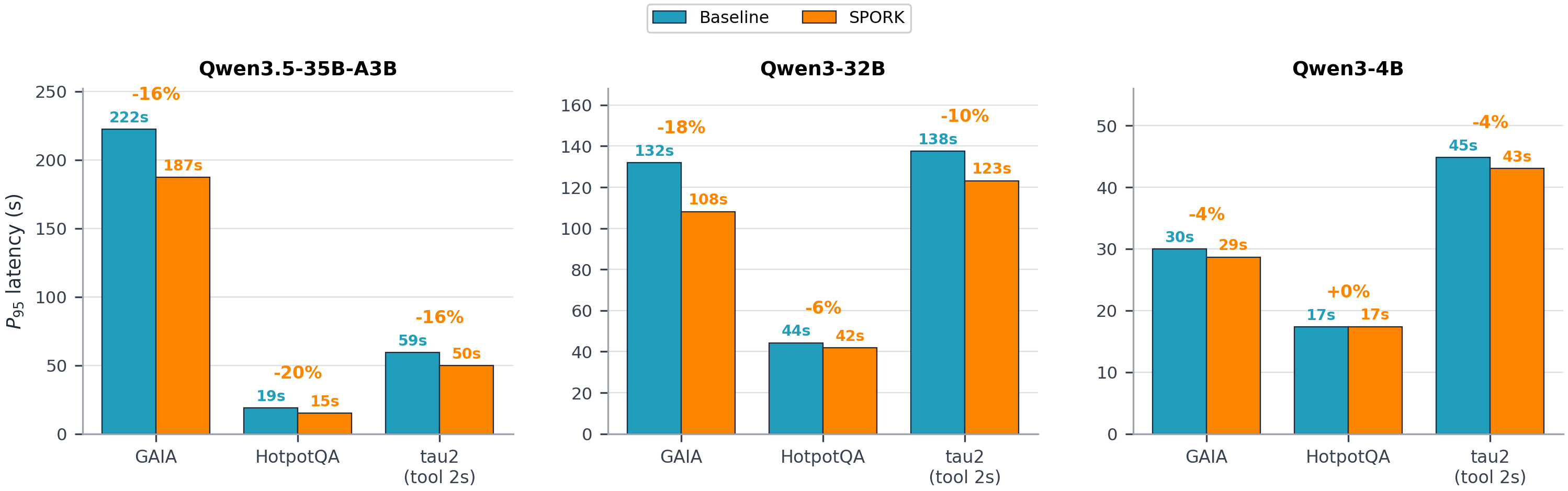}
  \caption{\textbf{\spork{} reduces $P_{95}$ tail latency across models and benchmarks.}
  Teal: baseline; orange: the paper-facing \spork{} configuration for each run
  (engine D1+D2+D3; HTTP D1+D2). $P_{95}$ latency on GAIA, HotpotQA, and
  tau2 (2\,s tool floor) for Qwen3.5-35B-A3B, Qwen3-32B, and Qwen3-4B (note the per-panel
  $y$-scales); the percentage above each pair is the $P_{95}$ reduction.
  \spork{} generalizes across model scale (4B$\to$32B) and architecture (dense and MoE).
  Per-model measurement setup is described in \S\ref{sec:setup}; quality is preserved
  across all configurations (Figure~\ref{fig:quality}).
  }
  \label{fig:hero_speedup}
\end{figure*}

\parabf{vLLM compatibility.}
\spork{}'s D1/D2 controller requires a serving backend that supports:
(1) streaming \texttt{/v1/chat/completions} with per-token logprobs;
(2) \texttt{/v1/completions} with per-token logprobs and an arbitrary prompt prefix
    (for the fork thread); and
(3) prefix KV-cache sharing between concurrent requests (D1 relies on this).
vLLM (our serving version, \S\ref{sec:setup}) satisfies all three with prefix caching
enabled; among the broader wave of serving engines~\citep{song_powerinfer_2024,
hu_deepserve_2025}, any backend exposing these three interfaces can host the controller.
D3 additionally requires access to the serving engine's speculative-draft proposer
interface; in our prototype this is implemented by wrapping vLLM's ngram proposer.

\parabf{Concurrency and safety.}
The main stream and the fork run as concurrent \texttt{asyncio} tasks against the same
vLLM endpoint.
A shared state dictionary (guarded by a \texttt{threading.Lock}) exposes the main
stream's growing chain-of-thought to each fork, so successive probes condition on more
context; the speculative tool runs in a worker thread via \texttt{asyncio.to\_thread}
while the main stream continues decoding.
Concurrent agent episodes, each contributing a main stream and a fork, run against vLLM
without observable throughput degradation (\S\ref{sec:experiments}).
By design, only tools marked read-only in a manifest are speculated.
Write and non-idempotent tools are not supported yet and always take the serial fallback
path; supporting them is feasible with sandboxing or agent-state checkpointing that can
roll back a mispredicted speculative call.
Our prototype evaluates read-only benchmarks (web search, page browse, Wikipedia,
DB-read), where the strict gate already bounds correctness on every turn.

\parabf{Deployment overhead.}
The controller adds $\leq 0.05$\,s of Python-level overhead per turn (thread
management, JSON parsing, lock acquisition) on top of the vLLM call latency.
This overhead is accounted for in $\Toh$ and is visible in the EQ1 validation
(\S\ref{sec:e2e_speedup}).

\parabf{Agent integration.}
In our agent harness, each turn wraps the stock ReAct-style loop: the main stream runs
through the chat API with tools registered in the request; the fork thread uses the same
prompt prefix plus a forced \texttt{<tool\_call>} opener and a JSON schema the model was
fine-tuned on.
Tool results are injected back into the chat history exactly as in the serial baseline, so
correctness reduces to matching the eventual tool name and arguments at turn end.
No changes to model weights or offline datasets are required: only serving configuration
(prefix caching on, logprobs enabled) and the controller process colocated with the agent.

\section{Evaluation}
\label{sec:experiments}

\subsection{Setup}
\label{sec:setup}

\paragraph{Models, Hardware, Frameworks.}
We evaluate \textbf{Qwen3-32B} and \textbf{Qwen3-4B} (dense)~\citep{yang_qwen3_2025} and
\textbf{Qwen3.5-35B-A3B} (a 3B-active mixture-of-experts)~\citep{qwen_qwen35_2026} to
test generalization across both scale and architecture, all self-hosted via
vLLM~0.19.1 (TP=1) on NVIDIA H20-3e GPUs (143\,GiB each) with bf16 precision.
All runs use greedy decoding (temperature~0, seed~42) with thinking mode enabled,
producing 2--20\,s of chain-of-thought decode that creates the overlap budget \spork{} exploits.

\paragraph{Benchmarks.}
Table~\ref{tab:benchmarks} (Appendix~\ref{app:configs}) profiles the three benchmarks.
GAIA and HotpotQA execute \emph{real} network tools (web search + page browse, and a
Wikipedia search API, respectively); tau2-bench executes structured DB-query tools with
simulated latency floors, giving a controlled sweep over $\Ttool$.

\paragraph{Metrics.}
\textbf{Latency}: per-query end-to-end wall-clock time.
\textbf{Quality}: exact-match accuracy (EM) for GAIA/HotpotQA, F1 for HotpotQA.
\textbf{Speedup}: $P_{95}$ ratio (baseline / \spork{}).

\paragraph{Configurations.}
Per-model serving mode, baseline, and \spork{} configuration appear in
Table~\ref{tab:configs} (Appendix~\ref{app:configs}); the dense-model baseline is
vLLM's built-in ngram speculative decoding~\citep{saxena_pld_2023}, and each model is
compared against its own baseline under the same serving mode.

\subsection{End-to-End Speedup}
\label{sec:e2e_speedup}

Figure~\ref{fig:hero_speedup} reports end-to-end $P_{95}$ latency across models and
benchmarks.
With Qwen3-32B, \spork{} achieves \textbf{$-$18\% $P_{95}$} on GAIA (131.9\,s $\to$
108.1\,s, real web search, $\Ttool = 4.8$\,s), while HotpotQA shows a modest
1.06$\times$ improvement because its short tools ($\Ttool = 0.9$\,s) sit near the EQ1
break-even.
The benefit transfers across model \emph{size}: Qwen3-4B achieves \textbf{1.15$\times$}
mean speedup on tau2-bench (simulated 2\,s tool floor), with full benchmark coverage in
Appendix~\ref{app:qwen4b}.
It also transfers across model \emph{architecture}: Qwen3.5-35B-A3B reduces $P_{95}$ by
16\% on tau2 (59.4\,s $\to$ 49.8\,s, N=155), 16\% on GAIA (222.4\,s $\to$ 187.3\,s,
N=53, tail-positive), and 20\% on HotpotQA (18.9\,s $\to$ 15.2\,s, N=200).

\begin{figure}[t]
  \centering
  \includegraphics[width=0.92\columnwidth]{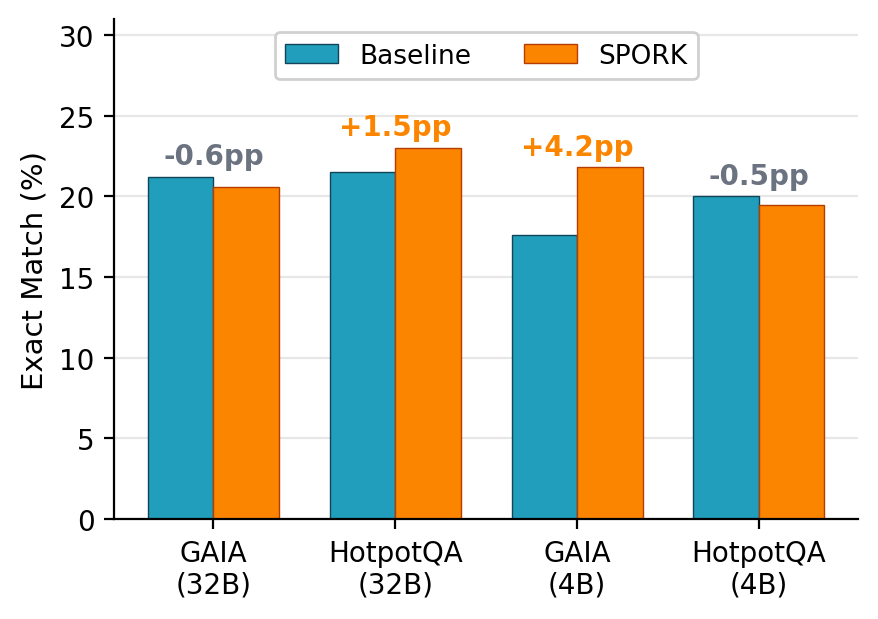}
  \caption{\textbf{\spork{} preserves task accuracy.} Paired comparison of baseline (teal)
  vs.\ \spork{} (orange) exact-match scores across benchmarks and model sizes.
  Quality stays within 1\,pp of baseline on all settings and sometimes improves
  (+1.5\,pp EM on HotpotQA with Qwen3-32B). The strict gate prevents speculative side
  effects from corrupting conversation history.}
  \label{fig:quality}
\end{figure}

\spork{} achieves $-$18\% $P_{95}$ on GAIA (Qwen3-32B) and 1.15$\times$
mean on tau2 (Qwen3-4B), confirming that the overlap window dominates when
tool latency $\geq$ 2\,s. \S\ref{sec:envelope} locates these points in the EQ1 envelope.

\subsection{Quality Preservation}
\label{sec:quality}

We expect \spork{} to preserve task accuracy: the strict gate ensures only
exact-matching tool calls are accepted, and D3 recycles only the rejected probe's
verified prefix as a speculative-decoding draft.
Speculative decoding's verification step preserves the target model's output
distribution~\citep{leviathan_fast_2023}; in \spork{} the target is the running model
itself, so D3 inherits this guarantee.

Figure~\ref{fig:quality} shows that \spork{} keeps EM within 1\,pp of baseline on
every benchmark and model, and sometimes improves it.
tau2-bench is not plotted because its quality is identical by construction: the
simulated DB tools return deterministic canned results, matching exactly for baseline
and \spork{}, so the latency sweep exercises timing only.
GAIA and HotpotQA instead call search APIs whose results are non-deterministic even
under identical arguments, so baseline and \spork{} trajectories cannot be
token-identical despite temperature~0; quality on these benchmarks is therefore
compared via aggregate scores (EM, F1) rather than per-token equality.
On HotpotQA with Qwen3-32B, \spork{} is faster
\emph{and} higher quality (+1.5\,pp EM, +0.012 F1) than the baseline.

\spork{} preserves accuracy (within $-$1\,pp of baseline) regardless of model size or
benchmark.
The strict gate is the key mechanism: it rejects any probe that does not exactly match
the main generation, ensuring the conversation history is never corrupted by speculative
side effects.

\subsection{Ablation}
\label{sec:ablation}

We expect each component (D1, D2, D3) to contribute independently to the final result,
with no single configuration dominating all metrics simultaneously.

\begin{figure}[t]
  \centering
  \includegraphics[width=0.92\columnwidth]{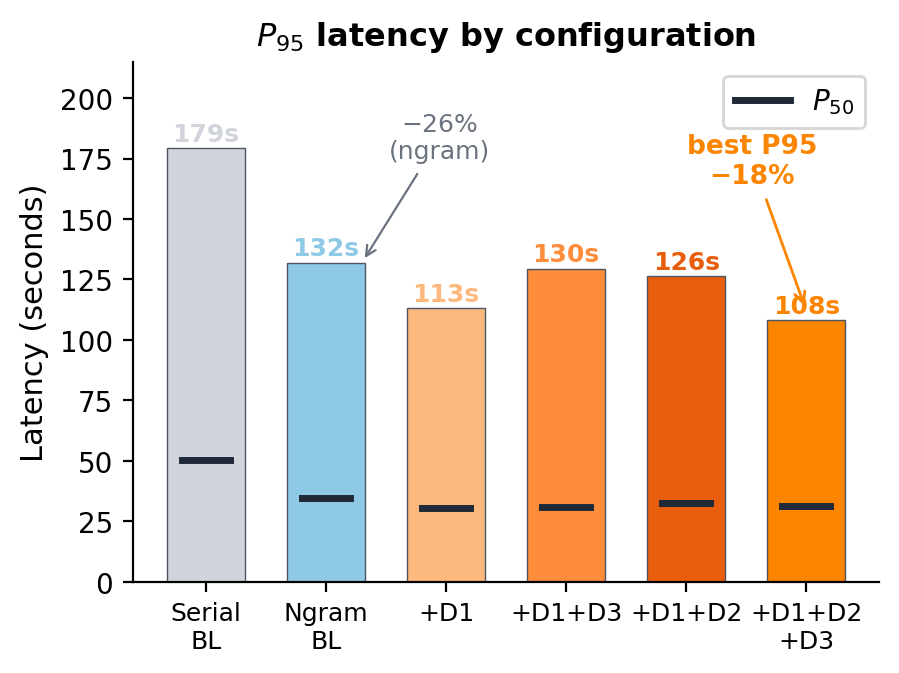}
  \caption{\textbf{Ablation: full \spork{} (D1+D2+D3) achieves the best tail latency.}
  GAIA N=165, Qwen3-32B; black dashes mark $P_{50}$.
  Ngram alone gives $-$31\% at $P_{50}$; D1+D2+D3 adds $-$18\% at $P_{95}$.
  No single config wins on all metrics: D1 for mean, D1+D2+D3 for $P_{95}$.}
  \label{fig:ablation}
\end{figure}

Figure~\ref{fig:ablation} decomposes contributions:
\begin{itemize}
  \item \textbf{Ngram spec-dec (the baseline)}: accelerates decode TPOT
    ($P_{50}$ 50.5$\to$34.7\,s, $-$31\%); every \spork{} configuration stacks on top of
    it, so the contributions below are orthogonal to token-level speculative decoding.
  \item \textbf{D1 on ngram}: additional $P_{50}$ $-$12\%, mean $-$10\%. Fork-prefix
    overlap adds on top of per-token spec-dec.
  \item \textbf{D1+D2+D3}: best $P_{95}$ (108.1\,s, $-$18\%). D2+D3 synergy helps
    tail queries where rejected retries leave verified prefixes that D3 recycles as
    drafts.
\end{itemize}

Per-configuration EM differences exist but stem from search-API and vLLM-batching
nondeterminism, both present in the baseline as well (\S\ref{sec:quality}); they do not
originate from the gate: D3 accepts only tokens that match the main model's greedy
output, so trajectory divergence is orthogonal to \spork{}'s correctness guarantee.

\parabf{D2 confidence gate analysis.}
D2 is a trade-off knob.
The gate raises per-turn acceptance from $\alpha \approx 0.22$ (D1, 49 accepted probes)
to $\alpha \approx 0.37$ (D1+D2, 101 accepted) and avoids wasted real tool calls by
filtering low-confidence probes before dispatch; the dispatched-probe totals (434 for
D1, 2124 for D1+D2) exceed the turn counts because the retry cadence re-probes
low-confidence turns (\S\ref{sec:eq1}).
The confidence threshold $\theta=0.90$ was selected by sweeping min-prob over 997 probes:
precision 88\%, recall 100\%, F1$=$0.937 (Figure~\ref{fig:d2_gate}).
This filtering pays when tools are slow or costly; when tools are cheap, the added wait
can outweigh the savings, and D2 alone can hurt mean latency on short-tool benchmarks.
With D3, the gate's rejected retries become useful drafts, and D2+D3 together give the
best $P_{95}$: the full configuration wins the tail while D1 alone wins the mean.

Each component contributes to different metrics: D1 for mean/P50, D2+D3 for $P_{95}$.
All three are needed for the best tail latency.

\subsection{Case Study: Self-Speculation vs.\ a Draft Model}
\label{sec:case_study}

\begin{figure}[t]
  \centering
  \includegraphics[width=0.92\columnwidth]{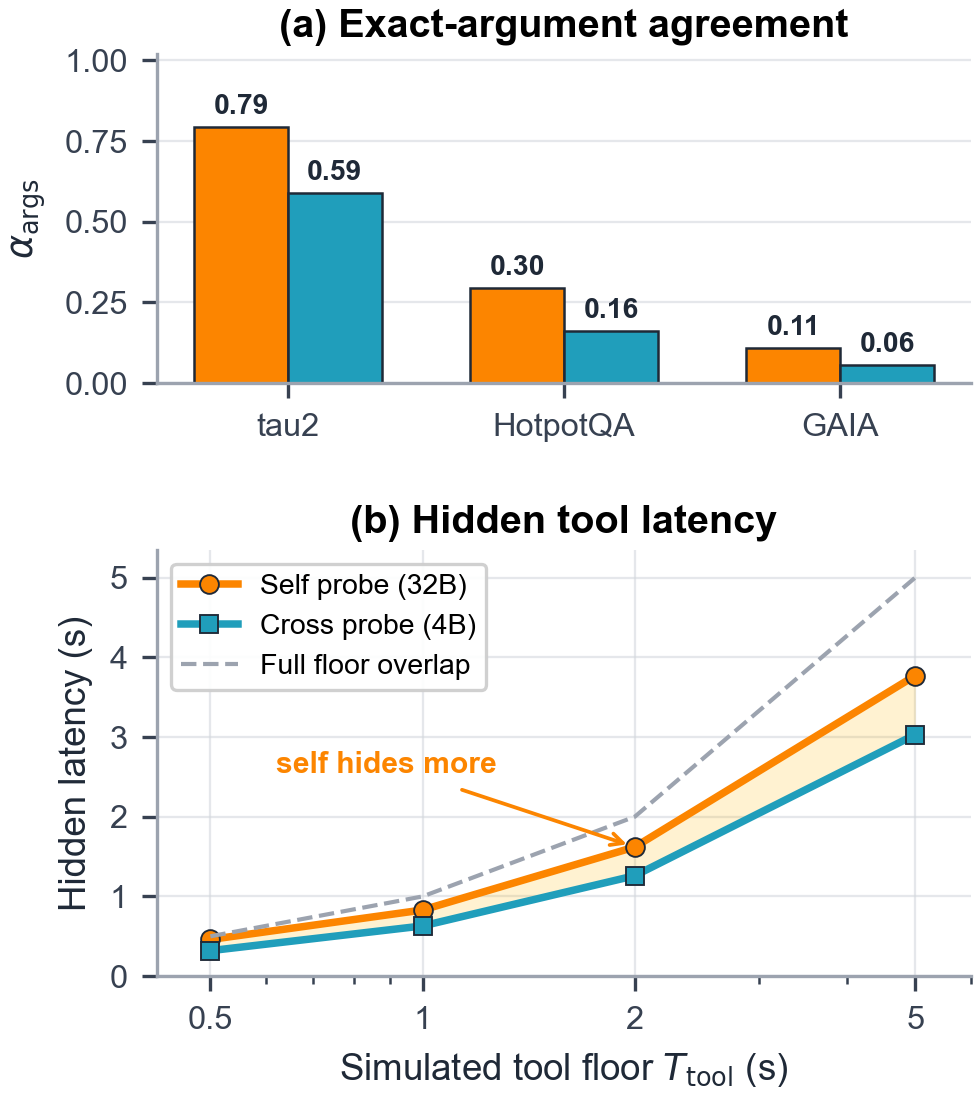}
  \caption{\textbf{Self-speculation trades a slower probe for higher useful overlap.}
  D1-only HTTP ablations with the same Qwen3-32B main model.
  \textit{(a)}~Self has higher exact-argument agreement than a Qwen3-4B cross probe
  on all three benchmarks.
  \textit{(b)}~On tau2, self hides more mean tool latency across floors; cross wastes
  2.1$\times$ more speculative tools at the 2\,s floor.}
  \label{fig:self_cross_ablation}
\end{figure}

A natural alternative to self-speculation is to serve a smaller model as a tool-call
drafter, analogous to token-level speculative decoding.
Figure~\ref{fig:self_cross_ablation} compares that
choice against self-speculation in a controlled D1-only HTTP ablation on tau2-bench:
same Qwen3-32B main model, tasks, seed, parser, and simulated tool floor; only the probe
model changes.

The cross-4B probe has lower probe latency, the time to decode the tool-call token
span: 0.315\,s mean vs.\ 0.970\,s for self at the tau2 2\,s floor.
Its exact-argument agreement, however, is far lower than self-speculation's: 0.590
vs.\ 0.792 on tau2 (Figure~\ref{fig:self_cross_ablation}\textit{a}), 0.161 vs.\ 0.296
on HotpotQA, and 0.056 vs.\ 0.109 on GAIA.
Since both probes usually finish before the tool floor on accepted turns, probe speed is
off the critical path; acceptance rate dominates realized overlap.
Self-speculation therefore hides more tool latency at every floor in the sweep and wastes
fewer speculative tool executions (154 vs.\ 317 at N=155;
Figure~\ref{fig:self_cross_ablation}\textit{b}).
It also avoids a second served model: in our measurement the cross-model arm serves two
models (Qwen3-32B and the Qwen3-4B drafter, each TP=1 on its own GPU), while
self-speculation needs only the target model's single GPU; a colocated drafter would
instead compete with the target for request scheduling and KV-cache capacity.
This result places a boundary on external drafters: they help only if their lower probe
cost offsets the lower $\alpha$ and the additional serving contention.

\subsection{Operating Envelope}
\label{sec:envelope}

We expect \spork{} to break even when tool latency exceeds probe overhead
($\Ttool \geq \Toh$, the probe overhead of \S\ref{sec:d1}) and the model produces
sufficient CoT for overlap.
Below this threshold, probe overhead dominates and \spork{} is neutral or negative.

\parabf{EQ1 calibration.}
The tau2 latency sweep validates EQ1's structural prediction: speedup grows monotonically
with tool latency (1.09$\times$ at 0.5\,s floor $\to$ 1.18$\times$ at 5.0\,s floor on
airline domain).
EQ1 predicts observed speedup with at most 1.84\% residual across all operating points
(1.84\% on tau2, 0.88\% on GAIA), confirming the cost model is a reliable predictor of
when \spork{} helps.

\begin{figure}[t]
  \centering
  \includegraphics[width=0.92\columnwidth]{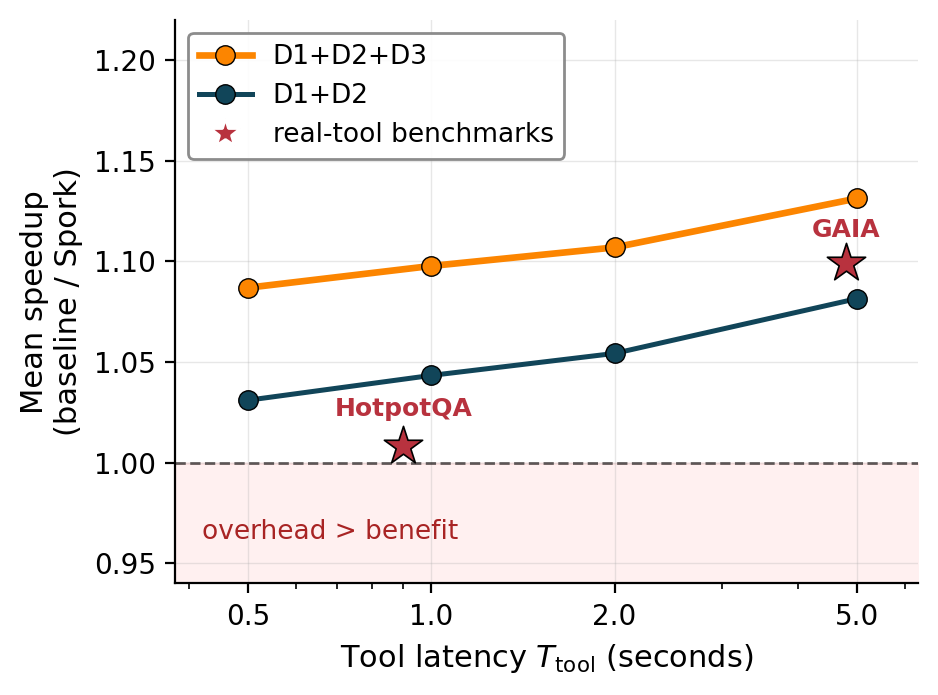}
  \caption{\textbf{\spork{} stays at or above break-even across the operating envelope.}
  Mean per-task speedup vs.\ tool latency on the tau2 sweep (N=155). Full \spork{}
  (D1+D2+D3) stays above break-even and rises with tool latency; the gated D1+D2 line
  approaches break-even on the shortest floors, matching EQ1's
  $\alpha\cdot\toverlap \geq \Toh$ condition.
  Stars mark the two real-tool operating points (Qwen3-32B mean speedup): near-neutral
  HotpotQA at EQ1's break-even, GAIA well inside the beneficial region.}
  \label{fig:operating_regime}
\end{figure}

Figure~\ref{fig:operating_regime} traces this envelope on the controlled tau2 sweep.
The real-tool benchmarks land where the cost model places them: GAIA's 4.8\,s tool
latency sits well inside the beneficial region, while HotpotQA's 0.9\,s latency sits
near break-even, exactly where EQ1 predicts the gains should shrink toward neutrality.

\parabf{When \spork{} wins.}
The largest gains appear when $\Ttool$ is large relative to probe overhead, tool schemas
are regular enough for high $\alpha$, and speculation is restricted to read-only
tools (\S\ref{sec:implementation}): GAIA's multi-second real web search is the
clearest favorable regime ($P_{95}$ $-$18\%).
On HotpotQA (0.9\,s Wikipedia API), \spork{} is Pareto-dominant with Qwen3-32B (faster and
higher EM) and Pareto-neutral with Qwen3-4B: it never hurts, and occasionally helps on
tail queries.
The case study (\S\ref{sec:case_study}) reinforces the principle: once the probe finishes
before the tool floor, acceptance rate, not probe speed, determines hidden tool latency.

\parabf{When \spork{} does not help.}
The primary boundary is a vanishing overlap window ($\toverlap \to 0$): from the decode
side, no-think mode cuts main decode to 0.3--0.9\,s, too fast for the probe to finish
before the main stream needs the tool result (0.79$\times$ on tau2), while thinking-mode
CoT, typically $\geq$2\,s in our workloads, creates the window; from the tool side, very
fast tools (e.g., local file reads) finish before any probe can pay off, leaving nothing
to hide.
The window also bounds long tools: $\toverlap$ ends when either side finishes, so tool
time that runs past the main stream's stop is not hidden, and the relative gain
saturates once $\Ttool$ exceeds the decode remaining at dispatch.
A second boundary, a native tool-call format that diverges from the forced probe
($\alpha \to 0$), is detailed in Appendix~\ref{app:format_divergence}.

\spork{}'s operating boundary is well-defined: thinking-mode CoT long enough to host the
probe, and tool latency above the probe-overhead break-even, with gains growing until
$\Ttool$ fills the available decode window.
Practitioners can evaluate applicability before full deployment with a lightweight pilot
trace measuring fork name accuracy, $\alpha$, and $\Ttool$, plugged into EQ1.

\FloatBarrier

\section{Related Work}
\label{sec:related}

\paragraph{Token-level speculative decoding.}
Speculative decoding~\citep{leviathan_fast_2023} accelerates generation with draft
models or extra heads that propose tokens the target model verifies in
parallel~\citep{li_eagle_2024, li_eagle2_2024, li_eagle3_2025, cai_medusa_2024,
an_pard_2025}; model-free variants draft from prompt or prior-output
n-grams~\citep{saxena_pld_2023, oliaro_suffixdecoding_2024}.
Both families shorten the \emph{decode} portion of a turn but leave the tool wait
untouched: the external call is still issued only after the model finishes emitting it.
\spork{} is orthogonal: our partial-token accept reuses vLLM's
speculative-decoding verification path to recycle a rejected probe's verified prefix as
draft tokens, so the two mechanisms stack.

\paragraph{Action- and execution-level speculation.}
Speculative Actions~\citep{ye_speculative_2026}, SPAgent~\citep{huang_reducing_2025},
DualSpec~\citep{zhong_dualspec_2026}, and PASTE~\citep{sui_act_2026} move speculation up to
the action level, pre-executing likely tool calls; \citet{nichols_optimizing_2025}
speculate tool calls inside the serving engine, targeting agent-serving throughput.
They differ from \spork{} in the \emph{source} of the prediction: Speculative Actions and
DualSpec rely on auxiliary predictor models or verifier policies,
\citet{nichols_optimizing_2025} draft with a smaller speculative model, and PASTE
mines recurring tool-call patterns from historical traces and replays them through a
middleware interceptor.
IdleSpec~\citep{choi_idlespec_2026} fills the same tool-wait windows with speculative
\emph{plans}, improving task accuracy rather than hiding tool latency.
\spork{} instead forks the running model's own state and reads its probe logprobs:
day-one speculation with no auxiliary model, no trace collection, and the prefix-cache
sharing an interceptor lacks.
Atomix~\citep{mohammadi_atomix_2026} adds complementary transactional semantics for
speculative side effects, which our read-only-tool restriction sidesteps.

\paragraph{Early intent in LLMs.}
Certaindex~\citep{fu_certaindex_2024} finds that reasoning programs' intermediate
answers stabilize well before generation ends and allocates test-time compute from
that certainty; When2Tool~\citep{sun_when2tool_2026} shows tool-call necessity is
linearly decodable from pre-generation hidden states.
DEER~\citep{yang_deer_2025} induces an early final answer to \emph{stop} reasoning, and
SpecExit~\citep{yang_specexit_2025} reads exit signals from draft-model hidden states;
inspired by this line of work, \spork{} inserts a forced tool-call prefix to surface
the next tool call early and overlap the tool wait with the remaining decode.

\section{Conclusion}
\label{sec:conclusion}

We presented \spork{}.
Our central finding is that agentic LLM workloads exhibit exploitable early intent: the
running model already knows its next tool call before finishing its reasoning, with
fork-at-start name accuracy of 74.6--99.6\% on Qwen3-32B against a tool wait of
16--37\% of agent wall time.
\spork{} turns this into latency with the lightest possible machinery, requiring no
draft model, no historical traces, and no retraining: D1--D3 map the early-intent
signals onto EQ1's terms, and our experiments validate the cost model on real web tools
(GAIA), short-latency APIs (HotpotQA), and the controlled tau2 latency sweep.
The mechanism generalizes across scale and architecture, from Qwen3-4B to Qwen3-32B and
a 3B-active mixture-of-experts, while preserving task accuracy.

\parabf{Limitations.}
\spork{} speculates only read-only tools today; supporting write operations is future
work, requiring sandboxing or agent-state checkpointing so a mispredicted speculative
call can be rolled back.
It relies on open-weight models served by an engine that exposes per-token logprobs and
a completion endpoint; closed-source APIs cannot host the fork thread.
Finally, we target latency-oriented serving with spare capacity; under heavy batching,
probe traffic competes with foreground decode and the overlap benefit shrinks.

\parabf{Future work.}
Sandboxed or checkpointed execution would extend speculation to write tools: a
mispredicted call rolls back to the pre-speculation agent state.
Deeper engine integration could schedule probes into idle batch slots, shrinking
$\Toh$ further and widening the favorable regime.

\bibliographystyle{abbrvnat}
\bibliography{literature}

\newpage
\appendix
\section{Full EQ1 Derivation}
\label{app:eq1}

Let a single agent turn have:
\begin{itemize}
  \item $\Tdec$: main generation wall time (post-prefill, includes CoT and tool-call decode)
  \item $\Ttool$: tool execution wall time
  \item $\Tbase = \Tdec + \Ttool$: serial baseline cost
\end{itemize}
Under \spork{}, let $\alpha$ be the gate acceptance rate and $\toverlap$ be the mean realized overlap on accepted turns.

On \textbf{accepted turns} ($p = \alpha$), the speculative tool execution overlaps $\toverlap$ seconds of tool time with decode:
\[
  T_{\mathrm{hit}} = \Tbase - \toverlap
\]
On \textbf{rejected turns} ($p = 1 - \alpha$), the agent falls back to serial execution with overhead $\Toh$, but prefix recovery (D3) lets the main stream skip re-decoding the recovered tool-call tokens, replacing $\Tbase$ with the realized base cost $\Tbasestar = \Tbase - T_{\mathrm{D3}}$ ($\Tbasestar = \Tbase$ without D3):
\[
  T_{\mathrm{miss}} = \Tbasestar + \Toh
\]
Expected per-turn cost:
\[
  \begin{aligned}
  \mathbb{E}[T] &= \alpha \cdot T_{\mathrm{hit}} + (1-\alpha) \cdot T_{\mathrm{miss}} \\
                &= \Tbase - \alpha \cdot \toverlap + (1-\alpha)(\Tbasestar - \Tbase + \Toh)
  \end{aligned}
\]
Latency ratio:
\[
  S = \frac{\Tbase}{\Tbase - \alpha \cdot \toverlap + (1-\alpha) \cdot (\Tbasestar - \Tbase + \Toh)}
\]
For notational simplicity, the main text uses $\Tbasestar$ in place of $\Tbase$ on the speculative path and $\Toh$ as a uniform overhead (applied regardless of accept/reject), giving the formula $\Tbase / (\Tbasestar - \alpha \cdot \toverlap + \Toh)$.
The floor-2s validation in \S\ref{sec:envelope} uses this simplified form.

\paragraph{Break-even condition.}
$S \geq 1$ requires $\alpha \cdot \toverlap \geq (1-\alpha)(\Tbasestar - \Tbase + \Toh)$; without D3 ($\Tbasestar = \Tbase$) this reduces to $\alpha \cdot \toverlap \geq (1-\alpha) \cdot \Toh$, or approximately $\alpha \cdot \toverlap \geq \Toh$ when $\Toh$ is treated as a fixed overhead.

\paragraph{Tool-fraction ceiling.}
When $\alpha = 1$ and $\toverlap = \Ttool$ (perfect overlap of all tool time):
\[
  S_{\max} = \frac{\Tdec + \Ttool}{\Tdec} = \frac{1}{1 - \ftool}
\]
where $\ftool = \Ttool / \Tbase$ is the tool-time fraction.
On BrowseComp ($\ftool = 0.366$): $S_{\max} = 1.58\times$.
The gap between $S_{\max}$ and realized speedup is explained by (a) $\toverlap < \Ttool$ (probe dispatch lag + turns with $\Ttool < T_{\mathrm{probe}}$) and (b) $\Toh > 0$ on strict mode.

\section{D3 Partial-Token Accept: Phase 0 Data}
\label{app:d3}

Offline analysis of 1268 probe--baseline pairs from BrowseComp-120, a 120-question
probe-analysis run (distinct from the 240-question end-to-end evaluation run
described in Appendix~\ref{app:browsecomp}).
Both arms decode under identical observations, so the prefix statistics are
unbiased:

\begin{center}
\small
\begin{tabular}{lc}
\toprule
Statistic & Value \\
\midrule
Mean first-mismatch position $k$ & 29.5 tokens \\
Median $k$ & 27 tokens \\
$P_{25}$ / $P_{75}$ & 17 / 35 tokens \\
\% pairs with $k \geq 10$ & 82\% \\
\% pairs with $k \geq 20$ & 61\% \\
\bottomrule
\end{tabular}
\end{center}

Even rejected probes share a substantial token prefix with the baseline.
Accepting $k = 29.5$ tokens and autoregressively decoding the remaining $50 - 29.5 = 20.5$ tokens reduces the decode time of a full ${\sim}50$-token tool call from $\sim 1.5$\,s to $\sim 0.9$\,s (verify pass + 20.5 tokens), saving $\sim 0.6$\,s on such rejected turns.
In this 120-question run, 52\% of strict-gate turns were rejected, giving a mean per-turn D3 saving of $\approx 0.31$\,s.

\section{BrowseComp Experimental Details}
\label{app:browsecomp}

\paragraph{Setup.}
Qwen3-32B via vLLM 0.18.1, TP=4 on 4$\times$ NVIDIA H20-3e (143\,GiB each), bf16, max\_model\_len=40960.
BrowseComp first 240 questions (indices 0--239, fixed seed).
Temperature 0, seed 42 on all calls.
MAX\_TURNS = 15. Workers = 4 (4 questions in flight).
Tools: a production web-search API plus a web-page crawl/browse API.

\paragraph{First-token timeout.}
3.4\% of turns (50/1487) had probe dispatch timeout (first-token threshold = 3s).
These are treated as aborted probes; the turn falls back to serial execution.
Not catastrophic for overall results but indicates that under workers=4 concurrency, probe latency occasionally exceeds the 3s window on long real-content prompts.
D2 (confidence gating with earlier abort) would reduce this failure mode.

\paragraph{Fork statistics.}
\begin{center}
\small
\begin{tabular}{lc}
\toprule
Metric & Value \\
\midrule
Total tool turns & 1487 \\
Name hit rate & 83.7\% (1245/1487) \\
Args-exact rate & 38.8\% (577/1487) \\
Spec used (strict) & 577 (38.8\%) \\
Spec wasted (strict reject) & 687 (46.2\%) \\
First-token timeouts & 50 (3.4\%) \\
\bottomrule
\end{tabular}
\end{center}

\section{BrowseComp Timing Decomposition}
\label{app:timing}

Wall-time decomposition of the April HTTP-mode runs of Appendix~\ref{app:browsecomp};
the later engine-mode rerun shows the same tool share (37\%) but a larger prefill share
(it lacks the HTTP path's shared prefix cache).

\begin{center}
\resizebox{\columnwidth}{!}{%
\begin{tabular}{lcc}
\toprule
Component & Fraction of $\Tbase$ & Mean (s) \\
\midrule
$f_{\mathrm{tool}}$ (tool execution) & 36.6\% & 8.92 \\
$f_{\mathrm{dec}}$ (decode: CoT + tool\_call) & 57.2\% & -- \\
\quad CoT decode & 54.6\% & -- \\
\quad Tool-call decode & 2.6\% & -- \\
Other (prefill, overhead) & 6.2\% & -- \\
\midrule
Realized $\toverlap$ (spec\_used turns) & -- & 1.03 \\
Probe dispatch latency (mean probe\_wall) & -- & 0.72 \\
\bottomrule
\end{tabular}%
}
\end{center}

The 36\% turns with $\Ttool < 0.72$\,s (< probe\_wall) have zero realizable overlap: the tool completes before the probe can even dispatch speculative execution.
The 47\% turns with $\Ttool \in [0.72, 2]$\,s have limited overlap ($\leq 1.28$\,s).
Only the 17\% with $\Ttool \geq 2$\,s achieve substantial overlap.
Hence mean $\toverlap = 1.03$\,s despite mean $\Ttool = 8.92$\,s: 83\% of turns fall in the first two buckets.

\section{Per-Model Serving Configurations}
\label{app:configs}

Table~\ref{tab:benchmarks} profiles the three benchmarks of \S\ref{sec:setup};
Table~\ref{tab:configs} lists the serving mode, baseline, and \spork{} configuration
for every model in \S\ref{sec:experiments}.

\begin{table}[h]
\caption{\textbf{Benchmark and tool profile.} Tool vocabulary size, typical tool
latency, and turn depth per benchmark.}
\label{tab:benchmarks}
\centering
\footnotesize
\begin{tabular}{@{}lrrll@{}}
\toprule
Benchmark & N & \#Tools & Typical $\Ttool$ & Turns \\
\midrule
GAIA       & 165 & 2 & 4.8\,s mean (real)      & 1--8 \\
HotpotQA   & 200 & 1 & 0.9\,s mean (real)      & multi-hop \\
tau2-bench & 155 & 30 (2 dom.) & 0.5--5\,s floors (sim.) & multi-turn \\
\bottomrule
\end{tabular}
\end{table}
The HTTP configuration drives a stock vLLM server through its OpenAI-compatible API
and supports D1+D2 only; the engine configuration adds the D3 proposer integration
inside vLLM (\S\ref{sec:d3}).
Qwen3.5-35B-A3B uses JSON-format tool prompting throughout, as its native XML
tool-call format is a self-speculation boundary case (\S\ref{sec:envelope}).

\begin{table}[h]
\caption{\textbf{Per-model serving mode, baseline, and \spork{} configuration.}}
\label{tab:configs}
\centering
\footnotesize
\begin{tabular}{@{}llll@{}}
\toprule
Model & Serving mode & Baseline & \spork{} \\
\midrule
Qwen3-32B & engine & ngram spec.\ dec. & D1+D2+D3 \\
Qwen3-4B\textsuperscript{\dag} & engine & ngram spec.\ dec. & D1+D2+D3 \\
Qwen3.5-35B-A3B\textsuperscript{\ddag} & HTTP (JSON) & serial & D1+D2 \\
\bottomrule
\end{tabular}

\smallskip
{\footnotesize \textsuperscript{\dag}\,The tau2 run uses HTTP: serial baseline, D1+D2.
\textsuperscript{\ddag}\,The GAIA run uses D1 only.}
\end{table}

All end-to-end runs use greedy decoding with a fixed seed (42).

\section{Format-Divergence Boundary}
\label{app:format_divergence}

A model whose native tool-call format diverges from the forced probe defeats
self-speculation by collapsing the acceptance term ($\alpha \to 0$).
Qwen3.5-35B-A3B's native XML predicts the \emph{end-goal} tool rather than the next
intermediate one, collapsing fork-at-start name accuracy to
$\alpha_{\mathrm{name}}{=}4.5\%$ despite 97\% parse success; this is a semantic
next-step mismatch, not a parser failure, and it motivates the JSON-format prompting
of Table~\ref{tab:configs}.

\section{Qwen3-4B Full-Benchmark Coverage}
\label{app:qwen4b}

\begin{figure}[t]
  \centering
  \includegraphics[width=\columnwidth]{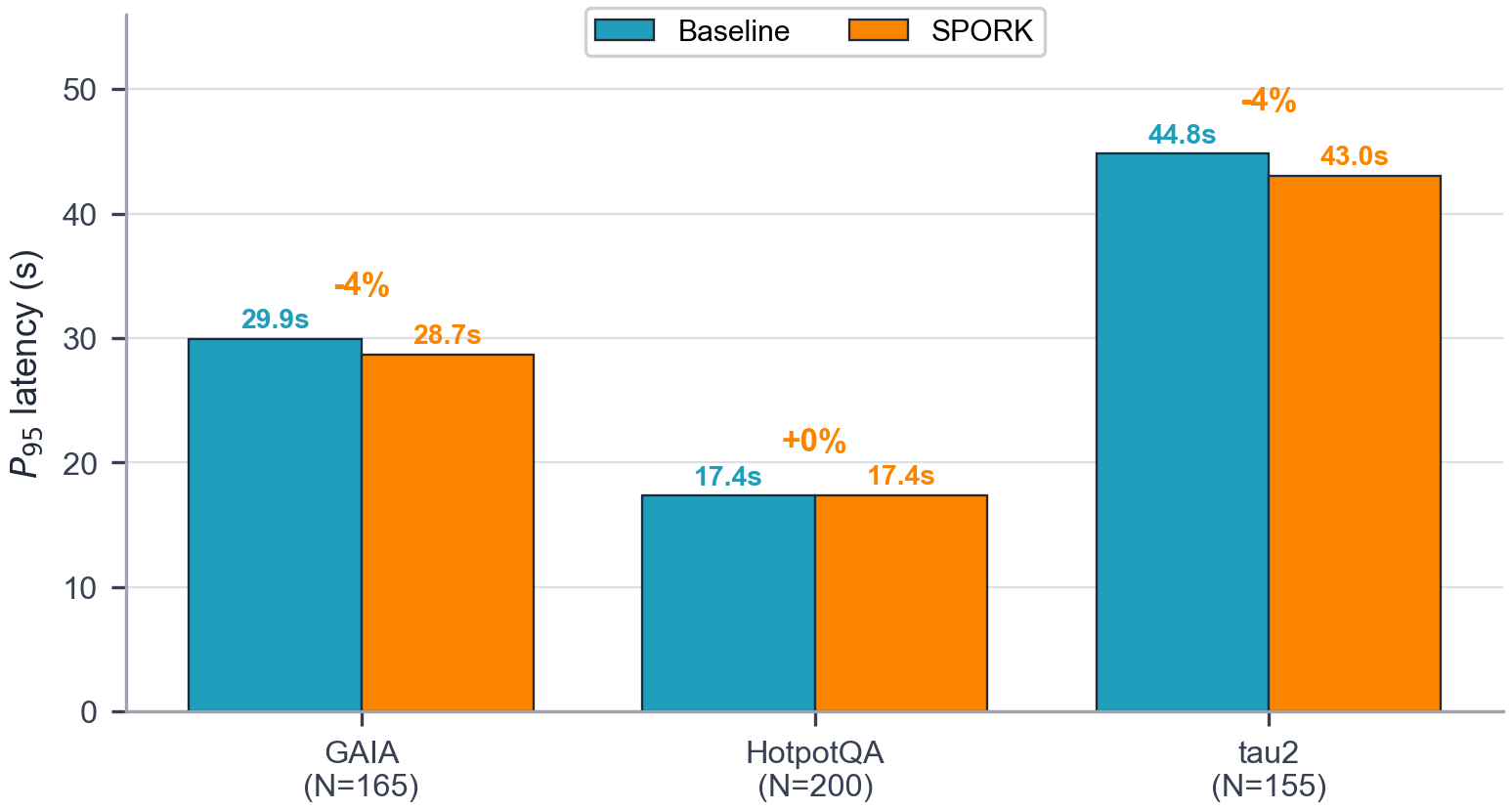}
  \caption{\textbf{Qwen3-4B $P_{95}$ latency across benchmarks (scaling boundary).}
  \spork{} vs.\ baseline (GAIA/HotpotQA: engine D1+D2+D3 vs.\ ngram spec-dec; tau2:
  HTTP D1+D2 vs.\ serial). tau2 sees a modest gain, GAIA is marginally positive,
  and HotpotQA is neutral. Quality is preserved or improved (GAIA EM $29\to36/165$;
  HotpotQA EM $40\to39/200$, within $-$1\,pp).}
  \label{fig:qwen4b_coverage}
\end{figure}

Figure~\ref{fig:qwen4b_coverage} and Table~\ref{tab:qwen4b_coverage} report the full
Qwen3-4B end-to-end coverage. These are
scaling-boundary diagnostics, not headline speedup claims: on the smaller model the
engine D1+D2+D3 protocol is near-neutral on HotpotQA and marginally positive on GAIA,
while the tau2 run (2\,s simulated tool-latency floor) shows a modest gain.
Even though Qwen3-4B in think mode produces long chains of thought (mean 3{,}323
CoT tokens on GAIA, 1{,}843 on HotpotQA), the smaller model decodes quickly, so the
overlap window and D3 recovery yield limited end-to-end benefit.

\begin{table}[t]
  \centering
  \small
  \begin{tabular}{lrrrl}
    \toprule
    Benchmark & N & Baseline $P_{95}$ & \spork{} $P_{95}$ & Speedup \\
    \midrule
    GAIA      & 165 & 29.95\,s & 28.67\,s & 1.03$\times$ (agg) \\
    HotpotQA  & 200 & 17.36\,s & 17.39\,s & 1.00$\times$ (agg) \\
    tau2      & 155 & 44.81\,s & 43.04\,s & 1.15$\times$ (mean)$^{\dagger}$ \\
    \bottomrule
  \end{tabular}
  \caption{Qwen3-4B end-to-end results. Latency columns are $P_{95}$; the GAIA and
  HotpotQA speedups (agg) are aggregate wall-clock ratios on real tools.
  $^{\dagger}$tau2 reports the mean speedup at the 2\,s simulated tool-latency floor
  (21.2$\to$18.5\,s), not a $P_{95}$ ratio.}
  \label{tab:qwen4b_coverage}
\end{table}

\section{Late-Probe Supersession (\texttt{continue\allowbreak\_after\allowbreak\_dispatch})}
\label{app:supersede}

D2 issues a probe after each additional block of CoT, and Insight~2 shows that probe
accuracy rises as the chain-of-thought unfolds. This raises a design choice: should the
controller commit to the \emph{first} probe that clears the gate, or keep probing
and upgrade its speculation when a later, higher-confidence probe predicts a
different call? We expose this as a single switch, \texttt{continue\allowbreak\_after\allowbreak\_dispatch}:
\begin{itemize}
  \item \textbf{OFF (default):} the first probe that clears the confidence gate dispatches
        the speculative tool and the retry loop stops (first commit wins).
  \item \textbf{ON:} the controller keeps probing after dispatch; if a later probe predicts
        a different call, it cancels the in-flight speculative tool and re-dispatches the
        new candidate (a \emph{supersession}).
\end{itemize}

Table~\ref{tab:supersede} reports a controlled A/B on tau2 (Qwen3-32B, HTTP mode, 2\,s tool
floor, $N{=}30$, a subset of the 155-task suite, seed~42) using deterministic record$\to$replay tool execution so that tool
latency and output are identical across arms (zero cache-miss fallbacks). Supersession is
mildly beneficial: it raises the gate acceptance rate ($\alpha$ $0.714\to0.762$) and roughly
halves wasted speculative dispatches ($28\to15$) through 36 supersessions, with no latency
regression (mean wall $80.1\to77.5$\,s, within noise).

\begin{table}[t]
  \centering
  \small
  \begin{tabular}{lrrr}
    \toprule
    Config & $\alpha$ (used/disp.) & wasted disp. & superseded \\
    \midrule
    baseline      & --             & 0  & 0  \\
    OFF (first commit) & 0.714      & 28 & 0  \\
    ON (supersede)     & 0.762 & 15 & 36 \\
    \bottomrule
  \end{tabular}
  \caption{Effect of \texttt{continue\allowbreak\_after\allowbreak\_dispatch} on tau2 (Qwen3-32B, HTTP, 2\,s floor,
  $N{=}30$, deterministic replay).}
  \label{tab:supersede}
\end{table}

The effect is small enough that we leave \texttt{continue\allowbreak\_after\allowbreak\_dispatch} as an optional
knob rather than a headline mechanism, and the main-paper results use the default behavior.
We do not report a GAIA latency A/B for this switch: with real free-text search, the main
stream's queries diverge across arms (no API seed under concurrent main/probe decode), so a
clean replay is not possible in HTTP mode; an engine-mode deterministic harness would be
required.

\end{document}